\documentclass[11pt]{article}
\usepackage{amsmath,amssymb,bbm,amsthm}
\usepackage{fullpage}
\usepackage{thm-restate,color,xcolor,xspace}
\usepackage{hyperref}
\hypersetup{colorlinks=true,%pdfborder={1 1 1 [3]},%
            citebordercolor={.6 .6 .6},linkbordercolor={.6 .6 .6},%
citecolor=blue,urlcolor=black,linkcolor=blue,pagecolor=black}
\usepackage[capitalise,nameinlink]{cleveref}

\makeatletter
\allowdisplaybreaks
\makeatother

\usepackage{algorithm}
\usepackage{algpseudocode}
\usepackage{algorithmicx}
\usepackage{setspace}
\usepackage[compact]{titlesec}
\allowdisplaybreaks

\newcommand{\f}{\frac}
\newcommand{\cd}{\cdot}
\newcommand{\bn}{\binom}
\newcommand{\sr}{\sqrt}
\newcommand{\cds}{\cdots}
\newcommand{\lds}{\ldots}

\newcommand{\sm}{\setminus}
\newcommand{\s}{\subseteq}

\newcommand{\BE}{\begin{enumerate}}
\newcommand{\EE}{\end{enumerate}}
\newcommand{\im}{\item}
\newcommand{\BI}{\begin{itemize}}
\newcommand{\EI}{\end{itemize}}

\newcommand{\Prod}{\displaystyle\prod\limits}

\newcommand{\R}{\mathbb R}

\newcommand{\N}{\mathbb N}
\newcommand{\Q}{\mathbb Q}

\newcommand{\de}{\delta}

\newcommand{\la}{\lambda}
\newcommand{\g}{\gamma}

\newcommand{\pt}{\partial}
\newcommand{\al}{\alpha}
\newcommand{\be}{\beta}
\newcommand{\om}{\omega}
\newcommand{\Om}{\Omega}
\newcommand{\el}{\ell}

\newcommand{\Th}{\Theta}
\newcommand{\m}{\mathcal}

\newcommand{\lf}{\lfloor}
\newcommand{\rf}{\rfloor}
\newcommand{\lc}{\lceil}
\newcommand{\rc}{\rceil}

\newcommand{\E}{\mathbb E}

\newcommand{\poly}{\text{poly}}

\newcommand{\calF}{\mathcal{F}}

\newcommand{\lp}{\left(}
\newcommand{\rp}{\right)}
\newcommand{\lb}{\left[}
\newcommand{\rb}{\right]}
\newcommand{\lmt}{\left[\begin{matrix}}
\newcommand{\rmt}{\end{matrix}\right]}

\newtheorem{theorem}{Theorem}
\newtheorem{lemma}[theorem]{Lemma}
\newtheorem{definition}[theorem]{Definition}
\newtheorem{corollary}[theorem]{Corollary}
\newtheorem{observation}[theorem]{Observation}
\newtheorem{claim}[theorem]{Claim}
\newtheorem{subclaim}{Subclaim}
\newtheorem{fact}[theorem]{Fact}
\newtheorem{assumption}[theorem]{Assumption}
\newcommand{\BT}{\begin{theorem}}
\newcommand{\ET}{\end{theorem}}
\newcommand{\BL}{\begin{lemma}}
\newcommand{\EL}{\end{lemma}}
\newcommand{\BD}{\begin{definition}}
\newcommand{\ED}{\end{definition}}
\newcommand{\BC}{\begin{corollary}}
\newcommand{\EC}{\end{corollary}}
\newcommand{\BO}{\begin{observation}}
\newcommand{\EO}{\end{observation}}
\newcommand{\BCL}{\begin{claim}}
\newcommand{\ECL}{\end{claim}}
\newcommand{\BSCL}{\begin{subclaim}}
\newcommand{\ESCL}{\end{subclaim}}
\newcommand{\BF}{\begin{fact}}
\newcommand{\EF}{\end{fact}}
\newcommand{\BA}{\begin{assumption}}
\newcommand{\EA}{\end{assumption}}
\newcommand{\BP}{\begin{proof}}
\newcommand{\EP}{\end{proof}}
\newcommand{\BPS}{\begin{proof}[Proof (Sketch)]}
\newcommand{\EPS}{\end{proof}}
\Crefname{observation}{Observation}{Observations}
\Crefname{claim}{Claim}{Claims}
\Crefname{subclaim}{Subclaim}{Subclaims}
\Crefname{fact}{Fact}{Facts}
\Crefname{assumption}{Assumption}{Assumptions}

\newcommand{\alert}[1]{{\color{red}#1}}

\newcommand{\tO}{\tilde{O}}

\newcommand{\thml}[1]{\label{thm:#1}}
\newcommand{\thm}[1]{\Cref{thm:#1}}
\newcommand{\leml}[1]{\label{lem:#1}}
\newcommand{\lem}[1]{\Cref{lem:#1}}

\newcommand{\clml}[1]{\label{clm:#1}}
\newcommand{\clm}[1]{\Cref{clm:#1}}
\newcommand{\corl}[1]{\label{cor:#1}}
\newcommand{\cor}[1]{\Cref{cor:#1}}

\newcommand{\eqnl}[1]{\label{eq:#1}}
\newcommand{\eqn}[1]{(\ref{eq:#1})}

\newcommand{\Venn}{\textup{Venn}}
\newcommand{\ex}{\textup{ex}}
\newcommand{\ol}{\overline}
\newcommand{\lk}{\overline\lambda_k}
\newcommand{\Fsmall}{\mathcal{F}_{\textup{small}}}

\renewcommand{\emptyset}{\varnothing}

\newcommand{\nn}{\widehat n}
\newcommand{\xx}{\widehat x}

\newcommand{\finalprob}{n^{-k} \cdot (k \ln n)^{-O(k^2 \ln \ln n)}}

\newcommand{\kcut}{\ensuremath{k\textsc{-Cut}}\xspace}

\newcommand{\kkclique}{\textsc{Max-Weight~\ensuremath{(k-1)}\textsc{-Clique}}\xspace}

\newcommand{\sun}{\mathsf{sf}}

\usepackage[colorinlistoftodos,textsize=tiny,textwidth=2cm,color=red!25!white,obeyFinal]{todonotes}

\newcommand{\initOneLiners}{%
    \setlength{\itemsep}{0pt}
    \setlength{\parsep }{0pt}
    \setlength{\topsep }{0pt}
}
\newenvironment{OneLiners}[1][\ensuremath{\bullet}]
    {\begin{list}
        {#1}
        {\initOneLiners}}
    {\end{list}}

\begin{document}

\title{\textbf{The Karger-Stein Algorithm is Optimal for $k$-cut}}
\author{ Anupam Gupta\thanks{{\tt anupamg@cs.cmu.edu}. Supported in part by NSF award CCF-1907820, and the Indo-US Joint Center for Algorithms Under Uncertainty. } \\ CMU \and Euiwoong Lee\thanks{{\tt euiwoong@cims.nyu.edu}. Supported in part by the Simons Collaboration on Algorithms and Geometry. }\\ NYU
  \and Jason Li\thanks{{\tt jmli@cs.cmu.edu}. Supported in part by NSF award
    CCF-1907820. }\\ CMU}
\date{}

\maketitle
\thispagestyle{empty}

\begin{abstract}
  In the $k$-cut problem, we are given an edge-weighted graph and want
  to find the least-weight set of edges whose deletion breaks the
  graph into $k$ connected components. Algorithms due to Karger-Stein
  and Thorup showed how to find such a minimum $k$-cut in time
  approximately $O(n^{2k-2})$. The best lower bounds come from
  conjectures about the solvability of the $k$-clique problem and a
  reduction from $k$-clique to $k$-cut, and show that solving $k$-cut
  is likely to require time $\Omega(n^k)$.  Our recent results have
  given special-purpose algorithms that solve the problem in time
  $n^{1.98k + O(1)}$, and ones that have better performance for
  special classes of graphs (e.g., for small integer weights).

  In this work, we resolve the problem for general graphs, by showing
  that for any fixed $k \geq 2$, 
  the Karger-Stein algorithm outputs any fixed minimum $k$-cut
  with probability at least $\widehat{O}(n^{-k})$, 
  where $\widehat{O}(\cdot)$ hides a $2^{O(\ln \ln n)^2}$ factor.
  This also gives an extremal bound of $\widehat{O}(n^k)$
  on the number of minimum $k$-cuts in an $n$-vertex graph
  and an algorithm to compute a minimum $k$-cut in similar runtime. 
  Both are tight up to $\widehat{O}(1)$ factors. 

  The first main ingredient in our result is a fine-grained analysis
  of how the graph shrinks---and how the average degree
  evolves---under the Karger-Stein process. The second ingredient is
  an extremal result bounding the number of cuts of size at most
  $(2-\delta) OPT/k$, using the Sunflower lemma.
\end{abstract}

\newpage

\setcounter{page}{1}

\section{Introduction}
\label{sec:introduction}

We consider the \kcut problem: given an edge-weighted graph
$G = (V,E,w)$ and an integer $k$, delete a minimum-weight set of edges
so that $G$ has at least $k$ connected components. This problem
generalizes the global min-cut problem, where the goal is to break the
graph into $k=2$ pieces. It was unclear that the problem admitted a
polynomial-time algorithm for fixed values of $k$, until the work of
Goldschmidt and Hochbaum, who gave a runtime of
$O(n^{(1/2 - o(1))k^2})$~\cite{GH94}. (Here and subsequently, the
$o(1)$ in the exponent indicates a quantity that goes to $0$ as $k$
increases.) The randomized minimum-cut algorithm of Karger and
Stein~\cite{KS96}, based on random edge contractions, can be used to
solve \kcut in $\tO(n^{2(k-1)})$ time. For deterministic algorithms,
there have been improvements to the Goldschmidt and Hochbaum
result~\cite{KYN06,Thorup08,chekuri2018lp}: notably, the tree-packing
result of Thorup~\cite{Thorup08} was sped up by Chekuri et
al.~\cite{chekuri2018lp} to run in $O(mn^{2k-3})$ time. Hence, until
recently, randomized and deterministic algorithms using very different
approaches achieved $O(n^{(2-o(1))k})$ bounds for the problem.

On the hardness side, one can  reduce $\kkclique$ to $\kcut$. It
is conjectured that solving \kkclique requires
$\tilde{\Omega}(n^{(1-o(1))k})$ time when weights are integers in the
range $[1,\Om(n^k)]$, and $\tilde{\Omega}(n^{(\omega/3)k})$ time for
unit weights; here  $\omega$ is the matrix multiplication constant.
Hence, these runtime lower bounds also extend to the $\kcut$,
suggesting that $\Omega(n^k)$ may be the optimal runtime for general
weighted $k$-cut instances.

There has been recent progress on this problem, showing the following
results:
\begin{enumerate}
\item We showed an $O(n^{(1.98 + o(1))k})$-time algorithm for general
  $\kcut$~\cite{GLL19}. This was based on giving an extremal bound on
  the maximum number of ``small'' cuts in the graph, and then using a
  bounded-depth search approach to guess the small cuts within the
  optimal $k$-cut and make progress. This was a proof-of-concept
  result, showing that the bound of $n^{(2-o(1))k}$ was not the right
  bound, but it does not seem feasible to improve that approach to
  exponents considerably below $2k$.
\item For graphs with polynomial integer weights, we showed how to
  solve the problem in time approximately
  $k^{O(k)} \, n^{(2\omega/3 + o(1))k}$~\cite{GLL18focs}. And for
  unweighted graphs we showed how to get the $k^{O(k)} n^{(1+o(1)k}$
  runtime~\cite{Li19focs}. Both these approaches were based on
  obtaining a spanning tree cut by a minimum $k$-cut in a small number of edges,
  and using involved dynamic programming methods on the tree to efficiently compute the edges and find the $k$-cut.
  The former relied on matrix multiplication ideas, and
  the latter on the Kawarabayashi-Thorup graph decomposition, both of
  which are intrinsically tied to graphs with small edge-weights.
\end{enumerate}

%\alert{We need to write some text about extending to general weights.}

In this paper, we %present what we think is the final word on the topic. We
show that the ``right'' algorithm, the original Karger-Stein algorithm,
achieves the ``right'' bound for general graphs. 
Our main result is the following.

\begin{restatable}[Main]{theorem}{Main}
\label{thm:main} Given a graph $G$ and a
parameter $k\geq 2$, the Karger-Stein
algorithm outputs any fixed minimum $k$-cut in $G$  with probability at least $\finalprob$.
\end{restatable}
For any fixed constant $k \geq 2$, the above bound becomes $n^k \cdot 2^{O(\ln \ln n)^2}$. 
This immediately implies the following two corollaries, where $\widehat{O}(\cdot)$ hides a quasi-logarithmic factor $2^{O(\ln \ln n)^2}$:

\BC[Number of Minimum $k$-cuts]\label{cor:enum} For any fixed $k \geq 2$, 
the number of minimum-weight $k$-cuts in a graph is at most $\widehat{O}(n^k)$.
\EC
This bound significantly improves the previous best bound $n^{(1.98 + o(1))k}$~\cite{GLL19}. 
It is also almost tight because the cycle on $n$ vertices has $\Omega(n^k)$ minimum $k$-cuts.

\BC[Faster Algorithm to Find a Minimum $k$-cut]\label{cor:runtime} 
For any fixed $k \geq 2$, there is a randomized algorithm that computes a minimum $k$-cut of a graph with high probability
in time $\widehat{O}(n^k)$.\footnote{While naively implementing the Karger-Stein algorithm takes time $O(n^2)$ per iteration, 
the well-known trick of reducing the number of vertices by a factor $1/2^{1/(2k-2)}$ and recursively running the algorithm twice
is known to reduce the total running time to $\widetilde{O}(n^{2k-2})$. Since our result is just an improved analysis of the same algorithm,
we can still apply the same trick to reduce the total running time to $\widehat{O}(n^{k})$. }
\EC
This improves the running time 
$n^{(1.98 + o(1))k}$ for the general weighted case~\cite{GLL19}
and even the running time $n^{(1+o(1))k}$ for the unweighted case~\cite{Li19focs}, where the extra $n^{o(k)}$ term is still at least polynomial for fixed $k$. 
It is also almost tight under the hypothesis that \kkclique requires $\tilde{\Omega}(n^{(1-o(1))k})$ time.

\subsection{Our Techniques}

Let us first recall the Karger-Stein algorithm:
\begin{algorithm}
    \caption{Karger-Stein Algorithm}
    \label{euclid}
    \begin{algorithmic}[1] % The number tells where the line numbering should start
        \Procedure{Karger-Stein}{$G = (V, E, w)$, $k \in \N$} \Comment{Compute a minimum $k$-cut of $G$}
            \While{$|V| > k$}% \Comment{We have the answer if r is 0}
                        \State Sample an edge $e \in E$ with probability proportional to $w(e)$.
                \State Contract two vertices in $e$ and remove self-loops. \Comment{$|V|$ decreases by exactly $1$}
            \EndWhile
            \State Return the $k$-cut of the original graph by expanding the vertices of $V$.
        \EndProcedure
    \end{algorithmic}
\end{algorithm}

In the spirit of \cite{GLL19}, our proof consists of two main parts:
(i)~a new algorithmic analysis (this time for the Karger-Stein
algorithm), and (ii)~a statement on the extremal number of ``small'' cuts
in a graph. In order to motivate the new analysis of Karger-Stein, let us first
state a crude version of our extremal result. Define $\la_k$ as the
minimum $k$-cut value of the graph. Think of $\lk:=\la_k/k$ % as the minimum $k$-cut divided by $k$,
as the average contribution of each of the $k$ components to the
$k$-cut. Loosely speaking, the extremal bound says the following:
\begin{quote}
  $(\star)$ There are a \emph{linear} number of cuts of a graph of
  size at most $1.99\lk$.
\end{quote}

To develop some intuition for this claim, we make two observations
about the cycle and clique graphs, two graphs where the number of
minimum $k$-cuts is indeed $\Omega(n^k)$.
%First, note that if $\lk$ were replaced by the minimum ($2$-)cut $\la$ in the graph, this statement would be false: the $n$-cycle has $\bn n2$ many cuts of size $\la=2$. 
Firstly, in the $n$-cycle, $\lk=1=\la/2$ for any value of $k$, and
since $1.99\lk=1.99<2=\la$, so there are \emph{no} cuts in the graph
with size at most $1.99\lk$, hence $(\star)$ holds. However, it breaks
if $1.99\lk$ is replaced by $2\lk$, since there are $\bn n2$ many cuts
of size $2\lk=2=\la$. Secondly, for the $n$-clique we have
$\la_k\approx k(n-1)$, since the minimum $k$-cut chops off $k-1$
singleton vertices. (We assume $k\ll n$, and ignore the $\bn k2$
double-counted edges for simplicity.) We have $\lk\approx n-1=\la$
instead for the $n$-clique, and there are exactly $n$ cuts of size at
most $1.99\lk$ (the singletons), so our bound $(\star)$ holds. And
again, $(\star)$ fails when $1.99\lk$ replaced by $2$. Therefore, in
both the cycle and the clique, the bound $1.99\lk$ is almost the best
possible. Moreover, the linear bound in the number of cuts is also
optimal in the clique. In general, it is instructive to consider the
cycle and clique as two opposite ends of the spectrum in the context
of graph cuts, since one graph has $n$ minimum cuts and the other has
$\bn n2$.

\subsubsection{Algorithmic Analysis.}
%\elnote{State K-S algo here, or in Intro? Fact that we can freely move between
%exponential clock and classical analysis?} 
To analyze Karger-Stein, we adopt an \emph{exponential clock} view of
the process: fix an infinitesimally small parameter $\de$, and on each
timestep of length $\de$, sample each edge with an independent
probability $\de/\lk$ and contract it. If $\de$ is small enough, then we can disregard
the event that more than one edge is sampled on a single
timestep. This perspective has a distinct advantage over the classical random contraction procedure that contracts one edge at a time: we can analyze whether each edge is contracted \emph{independently}. 
At the same time, we reemphasize it is just another view of the same process; 
conditioned on the number of vertices $n'$ in the remaining graph, the outcome of the exponential clock procedure has exactly the same distribution as the standard Karger-Stein procedure that iteratively contracted $n - n'$ edges. 

%Indeed our overall analysis adopts the exponential clock view until some time $T$, and finishes by using the classical Karger-Stein analysis. 
%However, it also comes at a cost, which we discuss later on.

Let us first translate the classical Karger-Stein analysis in the
exponential clock setting. Suppose that we run the process for
$2\ln n$ units of time (that is, $\f{2\ln n}\de$ timesteps of length
$\delta$ each). The probability that a fixed minimum $k$-cut remains at the end (i.e., has no edge contracted) is roughly
\[ \lp 1-\la_k \cd \f\de\lk \rp^{(2\ln n)/\de}\approx
  \exp\lp-\la_k\cd\f\de\lk\cd\f{2\ln n}\de\rp=\exp(2k\ln n)=
  n^{-2k}.\] 

How many vertices are there remaining after $2\ln n$ units of time? On
each $\de$-timestep, consider the graph before any edges are
contracted on this timestep. If there are currently $r\ge k-1$
vertices, then the sum of the degrees of the vertices must be at least
$r\lk$; otherwise, we can cut out $k-1$ random singletons to obtain a
$k$-cut of size at most $(k-1) \cd \lk<\la_k$ (which corresponds to a
$k$-cut of the same size in the original graph once we ``uncontract''
each edge), contradicting the definition of $\la_k$ as the minimum
$k$-cut. Therefore, the graph has at least $r\lk/2$ edges, which means
that we contract at least $r\lk/2 \cd \de/\lk=r\de/2$ edges in
expectation. We now claim that as $\de\to0$, this is essentially
equivalent to contracting at least $r\de/2$ \emph{vertices} in
expectation, since we should not contract more than one edge at any
timestep. That is, we contract a $\de/2$ \emph{fraction} of the
vertices per timestep, in expectation. Therefore, after $2\ln n$ units
of time (which is $\f{2\ln n}\de$ timesteps), assuming that there are
always at least $k-1$ vertices remaining, the expected number of
vertices remaining is at most
\[ n\cd \lp1-\f{\de}2\rp^{(2\ln n)/\de}\approx
  n\cd\exp\lp-\f{\de}2\cd\f{2\ln n}\de\rp= n\cd\f1n=1.\] Informally,
we should expect to be done at around time $2\ln n$. This argument is
not rigorous, since we cannot take a na\"{\i}ve union bound over the two successful events
(namely, that no edge in the fixed $k$-cut is contracted, and there
are $k-1$ vertices remaining at the end). We elaborate on how to
handle this issue later on.

At any given step of the Karger-Stein process, the bound of $r\lk/2$
edges can be tight in the worse case. So instead of improving the
analysis in a worst-case scenario, the main insight to our improvement
is a more average-case improvement. At a high level, as the process continues, we
expect more and more vertices in the contracted graph to have degree
\emph{much higher} than $\lk$. In fact, we show that the fraction of
vertices with degree at most $1.99\lk$ is expected to shrink
significantly throughout the process. Consequently, the total sum of
degrees becomes much larger than $r\lk$, from which we obtain an
improvement.

How do we obtain such a guarantee? Observe that every vertex of degree
at most $1.99\lk$ in the graph at some intermediate stage of the
process corresponds to a ($2$-)cut in the original graph of size at
most $1.99\lk$. Recall that by our extremal bound, we start off with
only $O_k(n)$ many such cuts. Some of these cuts can have size less
than $\lk$, but we show that there cannot be too many: at most
$2^{k-1}$ of them. The more interesting case is cuts of size in the
range $[\lk,1.99\lk]$: since each of these cuts has size at least
$\lk$, the probability that we contract an edge in a fixed cut is at
least $\lk \cd \de/\lk=\de$. This means that after just $\ln n$ units
of time (and not $2\ln n$), the cut remains intact only with
probability $(1-\de)^{(\ln n)/\de}\approx 1/n$. Taking an expectation
over all $O_k(n)$ cuts, we expect only $O_k(1)$ of them to remain
after $\ln n$ time, which means we expect only $O_k(1)$ many vertices
of degree at most $1.99\lk$ in the contracted graph after $\ln n$ time
has passed. This analysis works for any time $t$: we expect only
$O_k(n)\cdot (1-\de)^{t/\de}$ vertices of degree at most $1.99\lk$ in the
contracted graph after time $t$.

This upper bound on the number of small-degree vertices lower bounds
the number of edges in the graph, which in turn governs the rate at
which the graph shrinks throughout the process. To obtain the optimal
bounds, we model the expected rate of decrease as a differential
equation. In expectation, we find that by time $\ln n$ (not $2\ln n$),
we expect only $\widehat O_k(1)$ vertices remaining. This is perfect, since a
fixed minimum $k$-cut survives with probability $n^{-k}$ by this
time! To finish off the analysis from $\widehat O_k(1)$ vertices down to
$(k-1)$ vertices, we use the regular Karger-Stein analysis, picking up
an extra factor of $\lp\widehat  O_k(1)\rp^k=\widehat O_k(1)$.

Lastly, the issue of the $\widehat O_k(1)$ vertices bound holding only in
expectation requires some technical work to handle. In essence, we
strengthen the expectation statement to one with high probability. We
then union-bound over the event of $\widehat O_k(1)$ vertices remaining and the
event that the minimum $k$-cut survives---which only holds with
probability $n^{-k}$. This requires us to use concentration bounds
combined with a recursive approach; the details appear in
\S\ref{sec:conc} and \S\ref{sec:recur}.

\subsubsection{Extremal Result.}
\begin{figure}
\centering
\includegraphics[scale=.5]{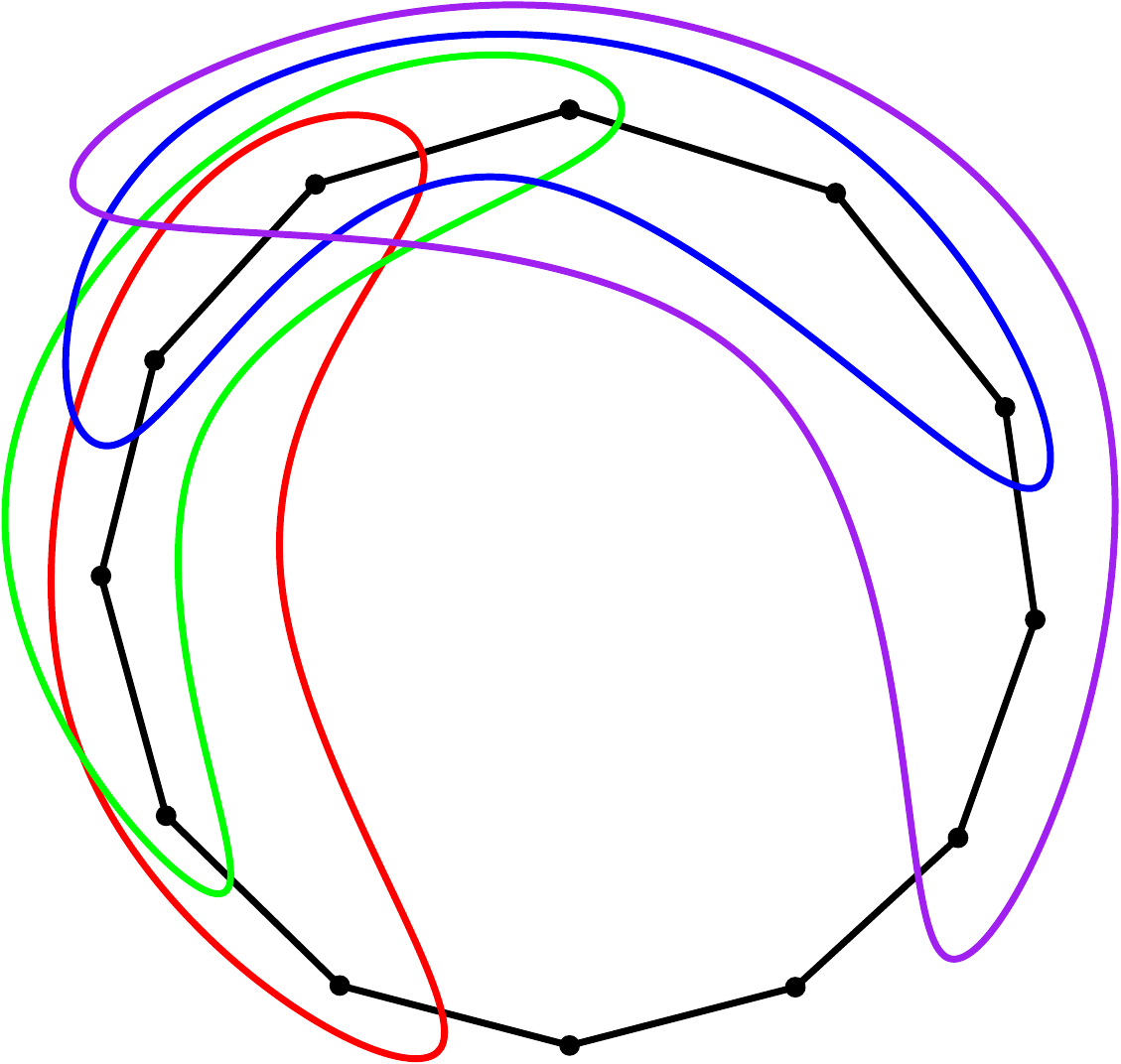}
\qquad
\includegraphics[scale=1]{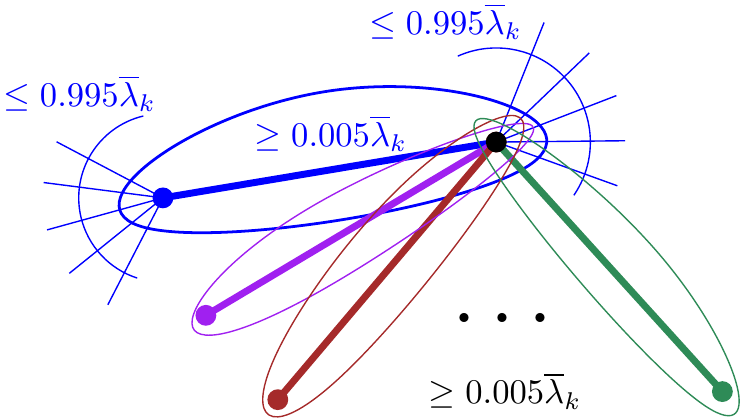}
\caption{\small \emph{Left: For a contradiction, suppose all $\bn n2$ cuts
  of the cycle have size at most $1.99\lk$. Then, we select $k/2=4$
  many such cuts as shown. Their Venn diagram has $8$ nonempty atoms
  and form an $8$-cut with cost
  $4 \cd 1.99\lk = 7.92\lk < 8\lk=\la_k$, contradicting the definition
  of $\la_k$ as the minimum $k$-cut. \quad Right: 
  A $401$-sunflower with nonempty core, with the core and all petals
  contracted to single vertices, each of degree at least $\lk$. Each
  bolded edge must have total weight at least $0.005\lk$ for
  the corresponding cut to have size at most $1.99\lk$. However, the
  $400$ many bolded edges excluding the blue one give total weight at
  least $400 \cd 0.005\lk=2\lk$, and each of them crosses the blue
  cut, contradicting the assumption that the blue cut has size at most
  $1.99\lk$.}}
\label{fig:1}
\end{figure}

Recall our target extremal statement $(\star)$: there are $O_k(n)$
many cuts of a graph of size at most $1.99\lk$.  Suppose for
contradiction that there are $\om_k(n)$ such cuts, and
assume for simplicity that $k$ is even. Our goal is to select $k/2$
of these cuts that ``cross in many ways'': namely, there are at
least $k$ nonempty regions in their Venn diagram (see
\Cref{fig:1}~left). This gives a $k$-cut with total cost
$k/2 \cd 1.99\lk < \la_k$, contradicting the definition of $\la_k$ as
the minimum $k$-cut.

To find such a collection of crossing cuts, one approach is to treat
each cut as a subset of vertices (the vertices on one side of the
cut), and tackle the problem from a purely extremal set-theory
perspective, ignoring the underlying structure of the graph. The
statement becomes: \emph{given a family of $\om_k(n)$ many distinct subsets
of $[n]$, there exist some $k/2$ subsets whose Venn diagram has at
least $k$ nonempty atoms.} In \cite{GLL19}, we tackled a similar
problem from this point of view.
% as a core technical component of our
% paper.
However, for our present problem, the corresponding extremal set
theory statement is too good to be true. The set system
$([n], \binom{[n]}{2})$ has $\bn n2$ sets, but no $k/2 = 3$ sets can
possibly form $k = 6$ regions. Hence, we need the additional structure of
cuts in a graph to formulate an extremal set theoretic statement that
holds.

Our key observation is that the cut structure of the graph  forbids
large \emph{sunflowers with nonempty core} in the corresponding set
family. To see why, consider a $401$-sunflower of sets
$S_1,S_2,\lds,S_{401}$ with nonempty core $C$, and suppose in addition
that the core $C$ and each petal $S_i\sm C$ is a cut of size at least
$\lk$. (Handling cuts of size less than $\lk$ is a technical detail,
so we omit it here.) For simplicity, consider contracting the core $C$
and petals $S_i\sm C$ into single vertices $c$ and $p_i$,
respectively. For each $i\in[401]$, the vertices $c$ and $p_i$ each
have degree at least $\lk$, and yet the cut $\{c,p_i\}$ has size at
most $1.99\lk$; a simple calculation shows that there must be at least
$0.005\lk$ edges between $c$ and $p_i$. Equivalently, in the weighted
case, the edge $(c,p_i)$ has weight at least $0.005\lk$ (see
\Cref{fig:1}~right). Now observe that the edges
$(c,p_2),(c,p_3),\lds,(c,p_{401})$ all cross the cut $\{c,p_1\}$, and
together, they have total weight $400\cd0.005\lk=\lk$. Hence, the cut
$\{c,p_1\}$ must have weight at least $2\lk$, contradicting the
assumption that it has size at most $1.99\lk$.

With this insight in mind, our modified extremal set theory statement
is as follows (when $k$ is even): given a family of $\om_k(n)$ many
distinct subsets of $[n]$ that \emph{do not contain a $401$-sunflower
  with nonempty core}, there exist some $k/2$ many subsets whose Venn
diagram has at least $k$ nonempty atoms. This statement turns out to
be true, and we provide a clean inductive argument that uses the
Sunflower Lemma as a base case. (Our actual extremal statement is
slightly different to handle the cuts of size less than $\lk$ as well
as odd $k$.)
% Also, we replace $1.99$ throughout with $(2-o(1))$ for a selectively
% chosen $o(1)$.)

\subsection{Preliminaries}
A weighted graph is denoted by $G = (V, E, w)$ where $w : E \to \Q^+$ gives positive rational weights on edges. 
Let $\lambda_k$ be the weight of a minimum $k$-cut, and $\lk := \lambda_k / k$. 
We use different logarithms to naturally present different parts of our analysis. We use $\ln$ to denote $\log_e$ and $\lg$ to denote $\log_2$.

%%% Local Variables:
%%% mode: latex
%%% TeX-master: "main"
%%% End:

\section{Random Graph Process}
\label{sec:random-graph-process}

In this section, we analyze the exponential clock viewpoint of the Karger-Stein algorithm to prove our main result \thm{main},
assuming the bound on the number of small cuts (\thm{ex-cuts}) proved in Section~\ref{sec:extremal-bounds}.

For the sake of exposition, throughout this section,
we assume $G$ is an unweighted multigraph, where the number of edges between a pair $(u, v)$ is 
proportional to its weight. 
Note that while it changes the number of edges $m$ and $\lk$, 
the resulting exponential clock procedure is still exactly equivalent to the standard Karger-Stein procedure in the original weighted graph,
and the final bounds in the main lemmas (\lem{expect} and \lem{conc}) only involve $n, k, t$.

\subsection{Expectation}
\label{sec:expectation}

In this section, we bound the number of vertices (in expectation) at
some point in the random contraction process:

\BL\leml{expect}
Suppose $G$ has at most $\be n$ many cuts with weight in the range $[\lk,\g \lk)$ for some constant $1\le\g<2$. Fix a parameter $t\ge0$, and suppose we contract every edge in $G$ with independent probability $1 - e^{-t/\lk}$. Then, the expected number of vertices in the contracted graph is at most $O(\f{\be+1}{2-\g})ne^{-(\g/2)t} + \f\g2(k-1)t$.
\EL

Consider the following \emph{exponential clock process}: let each edge $e$ of the graph independently sample a random variable $x(e)$ from an exponential distribution with mean $\lk$, which has c.d.f.\  $1-e^{-t/\lk}$ at value $t$. We say that the edge is \emph{sampled at time}  $x(e)$. Observe that for any $t\ge0$, every edge is sampled by time $t$ with probability exactly $1-e^{-t/\lk}$, so this process models exactly the one in the lemma. 

For a given $t\ge0$ and $\de>0$, the probability that an edge $e$ is sampled before time $t+\de$, \emph{given} that it is sampled after time $t$, is
\[ \f{\Pr[t \le x(e) \le t+\de]}{\Pr[t\le x(e)]} = \Pr[x(e) \le \de] = 1-e^{-\de/\lk} = \f\de\lk - O_{\lk}(\de^2) ,\]
where the first equality uses the \emph{memoryless} property of exponential random variables, and the $O_{\lk}(\cd)$ hides the dependence on $\lk$. Therefore, at the loss of the $O_{\lk}(\de^2)$ factor (which we will later show to be negligible), we can imagine the ``discretized'' process at a small timestep $\de$: the times $t$ are now integer multiples of $\de$, and each edge $e$ is sampled at the (discrete) time $t$ if $t \le e < t+\de$. Again, the probability that an edge is sampled at (discrete) time $t+\de$, given that it is sampled after time $t$, is $\de/\lk+O_{\lk}(\de ^2) \approx \de/\lk$. For technical purposes, we will not assume that $t$ is always a multiple of $\de$ in our formal argument. %In particular, if $\de\ll1/\poly(n)$, then we'll contract at most one edge per round, so it simulates the Karger-Stein random contraction process.% and for each $\de$-timestep, contract each edge of $G$ with probability $\de/\lk$. If $\de$ is small enough, then we'll contract at most one edge per round, so it simulates the Karger-Stein process.

Consider the graph at a given (discrete) time $t$, where we have contracted all edges sampled before time $t$ (but not those sampled at time $t$).
Suppose that there are $r=r(t)$ vertices in the contracted graph and $s=s(t)$ of them have degree in the range $[\lk,\g\lk)$. Note that at all times, at most $k-1$ vertices have degree~$<\lk$, since otherwise, we could take $(k-1)$ of those vertices (leaving at least one vertex remaining) and form a $k$-cut of weight less than $(k-1)\lk<\la_k$. Therefore, there are at least $r-s-(k-1)$ vertices with degree greater than $\g\lk$, so the number of edges is at least 
\[\f{s \cd \lk + (r-s-(k-1)) \cd \g\lk}2 = \f{\g r-\g s+s-\g(k-1)}2\lk = \lp\f\g2r - \f{\g-1}2s-\f\g2(k-1)\rp\lk.\]
For now, fix  $\de>0$, where $t$ is not necessarily an integer multiple of $\de$, and consider the time interval $[t,t+\de)$, where we contract all edges $e$ with $x(e)\in[t,t+\de)$. Each edge in the current contracted graph is contracted with probability $\de/\lk-O_{\lk}(\de^2)$ in this time interval, so we expect at least
\begin{gather}  \lp\f\g2r - \f{\g-1}2s-\f\g2(k-1)\rp\lk\cd\lp\f\de\lk -O_{\lk}(\de^2)\rp=  \lp\f\g2r - \f{\g-1}2s - \f\g2(k-1)\rp\de - O_{\lk,m,n}(\de^2)  \eqnl{edges-contr}
\end{gather}
edges to be contracted, where $O_{\lk,m,n}(\cd)$ hides dependence on $\lk,m,n$ (note that $r,s\le n$). Ideally, we now want to argue that every edge that is contracted in this interval reduces the number of remaining vertices by $1$. In general, this is not true if we contract a subset of edges that contain a cycle. However, if $\de$ is small enough (say, $\de\ll1/\poly(n)$), then in most cases, there is at most one edge contracted at all, in which case our desired argument holds. 

More formally, let $B_t$ be the (bad) event that more than one edge is contracted in the time interval $[t,t+\de)$. Then, by a union bound over all pairs of edges, we have 
\[\Pr[B_t] \le \bn m2\cd \lp\f{\de}{\lk}-O_{\lk}(\de^2) \rp^2 = O_{\lk,m,n}(\de^2) .\]
If the event $B_t$ holds, we will apply the trivial bound $\E[r(t+\de)]\le n$, and otherwise, we will use \eqn{edges-contr}. We obtain
\begin{align*}
 \E[r(t+\de)] &\le (1-\Pr[B_t])\cd\lp  r -  \lp\f\g2r - \f{\g-1}2s - \f\g2(k-1)\rp\de+ O_{\lk,m,n}(\de^2)\rp + \Pr[B_t] \cd n \\
&\le \lp r -  \lp\f\g2r - \f{\g-1}2s - \f\g2(k-1)\rp\de+ O_{\lk,m,n}(\de^2)\rp + O_{\lk,m,n}(\de^2) \cd n \\
&=  r -  \lp\f\g2r - \f{\g-1}2s - \f\g2(k-1)\rp\de+ O_{\lk,m,n}(\de^2).
\end{align*}
Taking the expectation at time $t$ and using linearity of expectation, we obtain
\begin{align*}
\E[r(t+\de)] &\le \E[r(t)] - \lp\f\g2\E[r(t)] - \f{\g-1}2\E[s(t)] - \f\g2(k-1)\rp\de+ O_{\lk,m,n}(\de^2).
\end{align*}
%\elnote{In discrete steps, (no. components decreased) can be different from (no. edges contracted). Shouldn't matter as $\delta \to 0$?}

We now bound $s=s(t)$ in terms of $t$. Every vertex whose degree is in the range $[\lk, \g\lk)$ must correspond to a cut in $G$ with weight in $[\lk,\g\lk)$, and by assumption, there are at most $\be n$ of them for some fixed constant $\be>0$. The probability that a cut of size $c\ge\lk$ has all its edges remaining up to time $t$ is $e^{-ct/\lk} \le e^{-t}$, so we expect at most $e^{-t} \cd \be n$ of these cuts to survive by time $t$. Therefore, $\E[s(t)]\le e^{-t}\be n$, and
\begin{align*}
\E[r(t+\de)] &\le \E[r(t)] - \lp\f\g2\E[r(t)] - \f{\g-1}2e^{-t}\be n - \f\g2(k-1)\rp\de + O_{\lk,m,n}(\de^2).
\end{align*}
Subtracting $\f\g2(k-1)(t+\de)$ from both sides, we obtain
\begin{align*}
\E\lb r(t+\de)-\f\g2(k-1)(t+\de)\rb &\le \E\lb r(t)-\f\g2(k-1)t\rb -  \lp\f\g2\E[r(t)] - \f{\g-1}2e^{-t}\be n\rp\de + O_{\lk,m,n}(\de^2).
%\\&\le  \E\lb r(t)-\f\g2(k-1)t\rb -  \lp\f\g2\E[r(t)] - \f{\g-1}2e^{-t}\be n\rp\de 
\end{align*}

We now solve for $\E[r(t)]$. Define $f(t):=\E[r(t)-\f\g2(k-1)t]$,
 so that
\begin{align*}
 f(t+\de)-f(t) &\le -  \lp\f\g2 \big[f(t) {+ \f\g2(k-1)t}\big] -
                 \f{\g-1}2e^{-t}\be n\rp\de+O_{\lk,m,n}(\de^2) \\
  &\le -  \lp\f\g2 f(t)  -
  \f{\g-1}2e^{-t}\be n\rp\de+O_{\lk,m,n}(\de^2).
\end{align*}
Taking $\de\to0$, we obtain the differential equation
\[ f'(t) = \lim_{\de\to0}\f{f(t+\de)-f(t)}\de \le -\f\g2f(t) + \f{\g-1}2e^{-t}\be n .\]
Set $B:=\f{\g-1}2\be$, so we instead have
\begin{gather}
 f'(t) \le -\f\g2f(t) + Be^{-t} n . \label{eq:f'}
 \end{gather}
Observe that if we had $f'(t)=-\f\g2f(t)$ instead, then that would solve to $f(t)\le e^{-(\g/2)t}n$, but there's the additional $Be^{-t}n$ term to deal with. However, $e^{-t}$ drops much faster than $e^{-(\g/2)t}$ (since $\g<2$ by assumption), so intuitively, the $Be^{-t}n$ factor doesn't affect us asymptotically. We now formalize our intuition.

To upper bound $f(t)$, we will solve the differential equation \eqn{f'} where we pretend the inequality in \eqn{f'} is actually an equality. More formally, define
\[ A:= \f{B}{B+1-\g/2}, \]
which satisfies $A<1$ since $\g<2$, and define
\[ \tilde f(t) := \f1{1-A}n(e^{-(\g/2)t} - Ae^{-t}). \]
%for some $A<1$ depending on $\g$ and $\be$, to be determined later. 
The following is a simple exercise in differential equations which we defer to the appendix.
%We will show that  $\tilde f(0)=f(0)$ and that  $\tilde f$ satisfies the differential equation \eqn{f'} with equality (where $f'$ is replaced by $\tilde f$), and therefore $\tilde f(t)\ge f(t)$ for all $t\ge0$.  

\BCL\clml{diffeq}
The function $\tilde f(t)$ satisfies $\tilde f(0)=f(0)$ and
\[ \tilde  f'(t) =-\f\g2\tilde f(t) + Be^{-t} n ,\]
which is the differential equation \eqn{f'} with equality (where $f$ is replaced by $\tilde f$). It follows that $\tilde f(t)\ge f(t)$ for all $t\ge0$.  
\ECL

Following \clm{diffeq}, we have
\begin{align*}
 \E\left[r(t)-\f\g2(k-1)t\right] = f(t) \le \tilde f(t) &\le \f1{1-A}n(e^{-(\g/2)t} - Ae^{-t}) \\&\le \f1{1-A}ne^{-(\g/2)t}
\\&= \f{B+1-\g/2}{1-\g/2} ne^{-(\g/2)t}
\\&= \f{\f{\g-1}2\be+1-\g/2}{1-\g/2} ne^{-(\g/2)t}
\\&= \lp\f{\g-1}{2-\g}\be+1\rp ne^{-(\g/2)t}
\\&=O\lp\f{\be+1}{2-\g}\rp ne^{-(\g/2)t} .
\end{align*}
Adding $\f\g2(k-1)t$ to each side finishes the proof of \Cref{lem:expect}.

% \medskip
% \alert{Jason: fixed it; does it look correct now? ---Could you guys please check: it seems that we want
%   \begin{align*}
%    \tilde f(t) &= \left( 1 + \frac{B}{1-\g/2} \right) n e^{-(\g/2)t} -
%     \frac{B}{1-\g/2} n e^{-t} \\
%     &= n e^{-(\g/2)t} + \frac{B}{1-\g/2} n(e^{-(\g/2)t} - e^{-t})  \\
%     &= n e^{-(\g/2)t} + \frac{\be(\g - 1)}{2-\g} n(e^{-(\g/2)t} - e^{-t}) .
%   \end{align*}
%   And then we should get $O(\frac{1+\be}{2-\g}) e^{-(\g/2)t}$, which
%   would be fine even when $\be = 0$.
% }

%%% Local Variables:
%%% mode: latex
%%% TeX-master: "main"
%%% End:

\subsection{Concentration}
\label{sec:conc}

In this section, we prove that for any graph with bounded number of edges, if we sample each edge independently with probability $p = 1 - e^{-t/\lk}$, the number of connected components is 
at most $\widetilde{O}(\sqrt{n})$ plus the expected value with high probability. 
It will be subsequently used in the recursive analysis in Section~\ref{sec:recur}.

%Message: additive $\sr n\log n$ concentration, we'll apply it with $\E[X] \approx n^{1/2}$ so concentration negligible

\BL\leml{conc}
Let $\al\ge1$, $t\ge\Om(1)$, and $N\ge n$ be parameters.
Let $G$ be a graph with at most $\al\lk n$ edges.
Suppose we sample every edge in $G$ with independent probability $1-e^{-t/\lk}$; let the random variable $f$ denote the number of connected components in the sampled graph. Then, with probability at least $1- N^{-2k}$, we have $f\le\E[f]+O(k \ln N \sqrt{\alpha t n})$.
\EL
\BP
Let $e_1, \dots, e_m$ be the edges of $G$, arbitrarily ordered.
For each $i \in [m]$, let $X_i \in \{ 0, 1 \}$ be the random variable indicating that $e_i$ is sampled. 
Then each $X_i$ is independent and $\Pr[X_i = 1] = p$ where $p = 1 - e^{-t / \lk}$. 
Let $f(X_1, \dots, X_m)$ be the number of components of the graph whose edge set is $\{ e_i : X_i = 1 \}$. 
For each $i \in [m]$, let 
\begin{align*}
& Y_i := \E [f(X_1, \dots, X_m) | X_1, \dots, X_{i}],  \\
& Z_i := Y_i - Y_{i - 1}, \\
& W_i := \sum_{j = 1}^{i} \E[Z_j^2 | X_1, \dots, X_{j - 1}].
\end{align*}
%For $m < i \leq N$, let $Y_i = f(X_1, \dots, X_m)$, $Z_i = W_i = 0$. 
Together with $Y_0 = \E[f]$, the sequence $\{ Y_0, \dots, Y_m \}$ forms a Doob martingale. 

Since the existence of one edge changes the number of connected components by at most $1$, $|Z_i| \leq 1$ always for every $i \in [m]$. 
For every $j \in [m]$ and $X_1, \dots, X_{j - 1}$ (which determines $Y_{j - 1}$), let $y_b := \E[f | X_1, \dots, X_{j - 1}, X_{j} = b]$ for $b \in \{0 ,1\}$. 
By the same argument, $|y_0 - y_1| \leq 1$, $y_1 \leq Y_{i - 1} \leq y_0$, and $Y_{i - 1} := p y_1 + (1 - p) y_0$, so that 
\[
\E[Z_j^2 | X_1, \dots, X_{j - 1}] = p \big( (1 - p) (y_1 - y_0) \big)^2 + (1 - p) \big( p (y_0 - y_1) \big)^2 \leq p(1-p).
\]
In particular, $W_i \leq pi$ for every $i \in [m]$ with probability $1$. 
We use the following concentration inequality for martingales, due to Freedman.

\BT\thml{freedman}\cite{freedman1975tail}
Let $\{ Y_0, \dots, Y_m \}$ be a martingale with associated differences
$Z_i := Y_i - Y_{i - 1}$, and 
\[ W_i := \sum_{j=1}^i \E[Z_i^2 \mid Y_1, \dots, Y_{k - 1}], \]
such that $|Z_i| \leq R$ and $W_i \leq \sigma^2$ for every $i$ with probability $1$. 
Then for all $t \geq 0$, 
\begin{equation}
\Pr[Y_m - Y_0 \geq s] \leq \exp\bigg( - \frac{s^2 / 2 }{\sigma^2 + Rs / 3} \bigg).
\eqnl{freedman}
\end{equation}
\ET
Plugging in $R = 1,\, \sigma^2 = pm,\, s = O(k \ln N \sqrt{\alpha t n})$ gives 
\[
\frac{s^2 / 2 }{\sigma^2 + Rs / 3} \geq 
\frac{s^2 / 2 }{pm + s / 3} \geq
\frac{s^2 / 2 }{\alpha t n   + s / 3} \geq
\Omega(\min(s^2 / (\alpha t n), s))
\geq \min(2 k^2 \ln^2 N, 2 k \ln N \sqrt{\alpha t n}) \geq 2 k \ln N.
\]
where the second inequality used the fact that $p = 1 - \exp(-t / \lk) \leq t / \lk$ and $m \leq \alpha \lk n$. 
Plugging in this bound to \eqn{freedman} proves the lemma.
\EP

How large do we have to set $\al$ in \lem{conc}? We show that $\al:=k$ suffices by first applying the graph sparsification routine of Nagamochi~and~Ibaraki, reducing its number of edges to at most $\la_k n$ while maintaining all minimum $k$-cuts.

\BT[Nagamochi-Ibaraki~\cite{NI92}]\label{thm:NI}
Given an unweighted graph $G$ and parameter $\la$, there exists a subgraph $H$ with at most $\la n  $ edges such that all $k$-cuts of size $\le\la$ are preserved. More formally, all sets $S$ with $|\pt_GS|\le\la$ satisfy $|\pt_GS|=|\pt_HS|$.
\ET
\BP
For $i=1,2,\lds,\la$, let $F_i$ be a maximal forest in $G\setminus\bigcup_{j<i}F_j$. For any edge $(u,v)$ in $G-H=G\setminus \bigcup_iF_i$, there must be an $(u,v)$ path in each $F_i$, otherwise we would have added edge $(u,v)$ to $F_i$. These $\la$ paths, along with edge $(u,v)$, imply that every cut that separates $u$ and $v$ has size~$\ge\la+1$.  Therefore, $u$ and $v$ must lie in the same component of any $k$-cut of size~$\le\la$, so removing edge $(u,v)$ cannot affect any such $k$-cut.
\EP

With \thm{NI} in hand, we now prove the following corollary which we will use in the next section, which combines \lem{conc} and the expectation statement of \lem{expect}.
\BC\corl{recursion-bound}
Let $t\ge\Om(1)$ and $N\ge n$ be parameters. Suppose $G$ has at most $\be n$ many cuts with weight in the range $[\lk,\g \lk)$ for some constant $1\le\g<2$.
Suppose we contract every edge in $G$ with independent probability $1-e^{-t/\lk}$. Then, with probability at least $1- N^{-2k}$, number of vertices in the contracted graph is at most $ O(\f{\be+1}{2-\g})ne^{-(\g/2)t} + \f\g2(k-1)t+O(k \ln N \sqrt{k t n})$.
\EC
\BP
First, apply \thm{NI} to the input graph $G$, obtaining a graph $H$ of at most $\la_kn$ edges with the same minimum $k$-cut value $\la_k$. We can imagine contracting the graph $G$ by first contracting each edge in $H$ with independent probability $1-e^{-t/\lk}$, and then contracting each edge in $G-H$ with the same probability. Applying \lem{expect} and \lem{conc} with $\al:=k$ on the graph $H$, we obtain that contracting the edges in $H$ alone gives us at most $ O(\f{\be+1}{2-\g})ne^{-(\g/2)t} + \f\g2(k-1)t+O(k \ln N \sqrt{k t n})$ with probability at least $1-N^{-2k}$. Contracting the edges in $G-H$ afterwards can only reduce the number of remaining vertices, so we are done. 
\EP

\iffalse
\subsection{Simplest Concentration?}
When $f : \{ 0, 1 \}^m \to \R$ is Lipshictz, McDiarmid gives the strongest concentration when each bit is unbiased. There must be a concentration inequality for Lipshictz functions on {\em biased domains}?
\fi

%%% Local Variables:
%%% mode: latex
%%% TeX-master: "main"
%%% End:

\subsection{Recursion}
\label{sec:recur}

In this section, we finish the proof of the main theorem, restated below:

\Main*

We will proceed by a recursive analysis:
\lem{expect} (expectation) and \lem{conc} (concentration), packaged together in \cor{recursion-bound}, show that if we let $t = \tfrac{1}{2}\ln n$ and contract each edge with probability $p = 1-e^{t/\lk}$, the number of remaining vertices becomes at most $\widetilde{O}(\sr n)$ with high probability. Also note that any fixed minimum $k$-cut $C$ survives (i.e., no edge in $C$ is contracted) with probability exactly $(1 - p)^{\lk} = n^{-k / 2}$. 

We then recursively call \Cref{cor:recursion-bound} on the contracted graph until the number of vertices becomes smaller than some threshold.
Formally, let $n_0: = n$ and $G_0 := G$. In the $i$th iteration, we set $t_i := \tfrac{1}{2}\ln n_{i-1}$ and contract each edge of $G_{i-1}$ with probability $p = 1-e^{t_i/\lk}$. 
The above analysis shows that with probability at least $(n_{i-1})^{-k / 2}$, no edge in $C$ is contracted and $n_i \leq \widetilde{O}(\sqrt{ n_{i-1}})$. 

If the second guarantee was precisely $n_i \leq \sqrt{n_{i-1}}$, iterating at most $T = \lg \lg n$ steps ensures that $n_T \leq O(1)$,
and the final probability that $C$ survives at the end is roughly at least $n^{-k/2}\cd n^{-k/4}\cd n^{-k/8}\cds \approx n^{-k}$. 
When the number of vertices becomes small, the naive Karger-Stein analysis can be applied. 
The proof below formalizes this intuition and accounts the fact that we can only ensure $n_i \leq \widetilde{O}(\sqrt{ n_{i-1}})$ in each iteration. 

\iffalse
 with $t=\f12 \ln \nn_{i-1}$, leaving at most $\tO(n^{1/4})$ vertices with good probability, and the minimum $k$-cut $C$ survives with probability roughly $n^{-k/4}$ on this step. This process continues on for roughly $\lg \ln n$ iterations, after which we expect $O_k(1)$ vertices to remain in the graph, and we will show that the minimum $k$-cut $C$ survives with probability roughly $n^{-k/2}\cd n^{-k/4}\cd n^{-k/8}\cds \approx n^{-k}$. Moreover, we prove that with probability roughly $n^{-k}$, the minimum $k$-cut $C$ survives \emph{and} there are $O_k(1)$ vertices remaining; this cannot be guaranteed by the expectation statement in \lem{expect} alone, since the probability $n^{-k}$ is too small to union bound over. Finally, with $O_k(1)$ vertices remaining, we can apply the naive Karger-Stein analysis on the remaining graph to show that $C$ is output with probability $O_k(1)^{-2k}$. Altogether, the Karger-Stein algorithm outputs $C$ with probability $\Om_k(1)\cd n^{-k}$. \alert{easiest way to argue that our recursive process simulates exactly Karger-Stein...?}
%We recursively subsample the contracted graph to eventually obtain a graph with $O_k(1)$ vertices, while proving that $C$ eventually survives with probability $\Omega_k(n^{-k})$. 
\fi

\begin{proof}
We prove the theorem by recursively applying \cor{recursion-bound} to reduce the number of vertices.
Given a graph $G$ with $n$ vertices, 
let $\gamma := 2 - 1 / \ln n$, and apply the following extremal theorem proved in Section~\ref{sec:extremal-bounds}.

\begin{restatable}[Extremal Theorem]{theorem}{ExtThm}
\thml{ex-cuts}
For any $\g<2$, there are at most $(\max(\frac{1}{2 - \gamma}, k))^{O(k)}n$ many cuts with weight less than $\g\lk$.
\end{restatable}

Let $\beta = (\max(\frac{1}{2 - \gamma}, k))^{O(k)} = (\max(\ln n, k))^{O(k)}$ so that there are at most $\beta n$ many cuts with weight less than $\g\lk$. 
The parameters $\beta$ and $\gamma$ will not change throughout the proof. 
Fix a minimum $k$-cut $C$ of $G$ so that $|C| = \lambda_k$. 

Let $n_0 = \nn_0 = n$ and $G_0 = G$. 
For each $i = 1, 2, \dots, T$ where $T = (\lg \lg n_0 - O(1))$, the $i$th iteration involves setting parameters $t_i := \frac{1}{2} \ln \nn_{i - 1}$, 
$\nn_i:=M \nn_{i-1}^{1/2}$
for some $M = O\lp\f{\be+1}{2-\g}\rp$ to be determined later, and contracting each edge in $G_{i - 1}$ with probability $p_i := 1 - e^{-t_i/\lk}$ to obtain $G_i$. 
Let $n_i$ be the number of vertices of $G_i$. 
In each iteration $i$, we want to ensure that the following events happen with high probability in each iteration, given that the same events happened in the previous iteration. 
\begin{OneLiners}
\item[1.] No edge in $C$ is contracted.
\item[2.] $n_i \leq \nn_i$.
\end{OneLiners}

For the event 1, the probability that no edge in $C$ is contracted is exactly 
\begin{align*}
(1 - p_i)^{\lambda_k} = e^{-k t_i} = (\nn_{i-1})^{-k/2}.
\end{align*}

For the event 2, we use \cor{recursion-bound} on $G_{i-1}$ with parameter $N\gets\nn_{i-1}$. 
Since $C$ is still a minimum $k$-cut of $G_{i-1}$, the minimum $k$-cut value of $G_{i-1}$ is still $\lk$. 
Applying \cor{recursion-bound} to $G_i$ with same $\beta$ and $\gamma$ ensures that with probability at least
$1-(\nn_{i-1})^{-2k}$, % , \eqnl{prob}
$n_i$ is at most 
\[ O\lp\f{\be+1}{2-\g}\rp n_{i-1} \exp\lp-\f\g2 t_i \rp+\f\g2 (k-1)t_i + O\lp k \ln \nn_{i-1} \sqrt{k t \nn_{i-1}} \rp  .\]
The last term in the above expression is at most $O\lp\f{\be+1}{2-\g}\rp (\nn_{i - 1})^{1/2}$ using the fact that $(k \ln \nn_{i-1} \sqrt{k t}) \leq O((\beta + 1) / (2 - \gamma))$. The first three terms can be upper bounded by
\begin{align*}
& O\lp\f{\be+1}{2-\g}\rp n_{i-1} \exp\lp-\f\g2 t_i \rp+\f\g2 (k-1)t_i  \\
\leq \quad &  O\lp\f{\be+1}{2-\g}\rp n_{i-1} \exp\lp-\f\g2 t_i \rp && \lp \frac{\gamma}{2}(k - 1)t_i \leq k \ln n_0 \leq O\lp \frac{\beta + 1}{2 - \gamma} \rp \rp \\
\leq\quad&  O\lp\f{\be+1}{2-\g}\rp n_{i-1} (\nn_{i-1})^{- \gamma / 4}  && (\mbox{definition of }t_i)  \\
\leq\quad&  O\lp\f{\be+1}{2-\g}\rp (\nn_{i - 1})^{1/2} && ((\nn_{i-1})^{- \gamma / 4} = (\nn_{i-1})^{- 1/2 + 1/4(\ln n_0)} = O(\nn_{i-1})).
\end{align*}
%Let $t:=\f12\ln n$, and contract each edge in $G$ with probability $1 - e^{-t/\lk}$. By \lem{expect}, the expected number of vertices $\E[X]$ in the contracted graph is at most \elnote{logs: default is ln and used lg if necessary}
%\elnote{Write proof of NI-type lemma? }
%Also by \lem{conc} (using $\alpha \leftarrow k$ and $N \leftarrow \nn_{i - 1}$), 
%with probability at least
%$1-(\nn_{i-1})^{-2k}$, % , \eqnl{prob}
%$n_i$ is at most 
%\begin{gather}
%\E[\nn_i] + O\lp k \ln \nn_{i-1} \sqrt{k t \nn_{i-1}} \rp \le 
%O\lp\f{\be+1}{2-\g}\rp (\nn_{i - 1})^{1/2} =: M (\nn_{i - 1})^{1/2},
%\eqnl{ni}
%\end{gather}
%again using the fact that $(k \ln \nn_{i-1} \sqrt{k t}) \leq O((\beta + 1) / (2 - \gamma))$.
It follows that with probability at least $1-(\nn_{i-1})^{-2k}$, we have 
\begin{gather}
\nn_i\le M (\nn_{i - 1})^{1/2}\eqnl{ni}
\end{gather}
 for large enough $M=O\lp\f{\be+1}{2-\g}\rp $.

Taking a union bound, the probability that events 1 and 2 both happen is at least 
\[
(\nn_{i-1})^{-k/2} - (\nn_{i-1})^{-2k} = 
(\nn_{i-1})^{-k/2} \cdot ( 1 - (\nn_{i-1})^{-3k / 2}).
\]

Let $\xx_i:=\lg \nn_i$ for all $i\ge0$, so that from \eqn{ni}, we obtain
\[
\xx_{i} \le \lg M +\f12\xx_{i-1},  
\]
which implies that 
\begin{gather}
\xx_i \le \lg M \cdot \lp 1 + \frac{1}{2} + \frac{1}{4} + \dots + 2^{-i+1} \rp + 2^{-i} \xx_0 \leq 2 \lg M + 2^{-i} \xx_0,
\eqnl{xi}
\end{gather}
so with $T := \lg \xx_0 - O(1)  = \lg \lg n_0 - O(1)$ steps, 
$\xx_T = O(\lg M)$, which translates to $\nn_T = M^{O(1)}$. 

We finally compute the probability that both events happen for each $i = 1, \dots, T = \lg \lg n_0 - O(1)$.
\begin{equation}
\prod_{i=1}^T \lp (\nn_{i-1})^{-k/2} \cdot ( 1 - (\nn_{i-1})^{-3k / 2}) \rp
= 
\lp \prod_{i=1}^T (\nn_{i-1})^{-k/2} \rp \cdot \lp \prod_{i=1}^T ( 1 - (\nn_{i-1})^{-3k / 2}) \rp 
\label{eq:prob}
\end{equation}

For the second product, the recursive definition $\nn_i:=M \nn_{i-1}^{1/2}$ also implies $\nn_i \geq n^{2^{-i}}$, 
so we can choose $O(1)$ in the definition $T = \lg \lg n_0 - O(1)$ to ensure $\nn_{T - 1} \geq 5$. 
Then the second product can be shown to be at least $\Omega(1)$, as 
\[
\lp \prod_{i=1}^T ( 1 - (\nn_{i-1})^{-3k / 2}) \rp \geq 1 - \sum_{i = 1}^T (\nn_{i-1})^{-3k/2},
\]
and the sequence $\{ (\nn_i)^{-3k / 2} \}$ is at least exponentially increasing with the last term at most $5^{-3}$ (since $k \geq 2$).

For the first product of~\eqref{eq:prob}, we use the fact $\xx_i \leq 2 \lg M + 2^{-i} \xx_0$ to bound 
\begin{align*}
\sum_{i=1}^T \xx_{i - 1} \leq 2T\lg M + 2\xx_0,
\end{align*}
which leads to 
\[
\lp \prod_{i=1}^T (\nn_{i-1})^{-k/2} \rp 
= \lp \prod_{i=1}^T (\nn_{i-1}) \rp^{-k/2}
\leq 2^{\lp \sum_{i=1}^T (\xx_{i-1}) \rp \cdot (-k/2)}
= \lp n^2 M^{2T} \rp^{-k/2} = n^{-k} \cdot M^{-Tk}. 
\]

Therefore, with probability at least $\Omega(n^{-k} \cdot M^{-Tk})$, events 1 and 2 happen for $i = 1, \dots, T$ which means that no edge in $C$ is contracted and $n_T \leq M^{O(1)}$. 
After this point, we can switch the standard Karger-Stein analysis of the same process where exactly one edge is contracted in each iteration. 
It shows that if $C$ will be output with at least $M^{-O(k)}$. Altogether, the minimum $k$-cut $C$ survives with probability at least (using $M \leq 
(\max(\ln n, k))^{O(k)} \cdot \ln n \leq 
(k \ln n)^{O(k)}$ and $T \leq \lg \lg n$), 
\[
n^{-k} \cdot M^{-O(Tk)} = n^{-k} \cdot (k \ln n)^{-O(k^2 \ln \ln n)}. 
\]

This completes the proof.
\end{proof}

%\alert{Apply recursively, to get rid of the $\logn$ factor? Need to bound \eqn{xi} more efficiently; the $4\log(2\xx_0)$ factor in \eqn{ind} is what's producing the $\logn$ factor right now.}

%For $t:=\f2\g\ln n$, we have $f(t) \le \tilde f(t) \le \f1{1-A}$, so $t=\f2\g\ln n+\f2\g\ln\f\be{2-\g}+O(1)$ suffices to obtain $f(t)\le1$, which means at most $\f\g2(k-1)+1$ components remaining.

%From the argument before, the probability that an $\al$-mincut survives is roughly
%\[ e^{-\al t} = \exp\lp-\al\lp\f2\g\ln n+\f2\g\ln\f\be{2-\g}+O(1)\rp\rp = O(\f\be{2-\g}n)^{-(2/\g)\al} .\] 
%This isn't completely formal, since everything's in expectation, but hopefully that's not too annoying...

%We will plug in $\g:=2-\f1\logn$, so that $2/\g=1+O(1/\logn)$. By \thm{ex-cuts} on this value of $\g$, there are $(k\logn)^{O(k)}n$ many cuts with weight less than $\g\lk$. Then, by \thm{ks} with $\be:=(k\logn)^{O(k)}$ and $\al:=k$, the number of cuts of size at most $\la_k=k\lk$ is at most
%\[ O\lp\f\be{2-\g}n\rp^{(2/\g)\al} = (k\logn)^{O(k)}n^{(1+O(1/\logn))k} = (k\logn)^{O(k)}n^{k} .\]

%%% Local Variables:
%%% mode: latex
%%% TeX-master: "main"
%%% End:

\section{Cuts and Sunflowers}
\label{sec:extremal-bounds}

In this section, we prove that 
for any $\gamma < 2$,
every graph has a small number of cuts whose weight is $\gamma \lk$.
Our main result in this section is:

\ExtThm*

\subsection{The Sunflower Lemma, and Refinements}
\label{sec:sunflowers}

Recall that given a set system $\m F$ over a universe $U$, an {\em
  $r$-sunflower} is a collection of $r$ subsets $F_1, \dots, F_r \in \m
F$ such that their pairwise intersection is the same:
there exists a {\em core} $S \subseteq U$ such that $F_i \cap F_j = S$
for all $i, j$, and hence $\cap_i F_i = S$.
Let $\sun(d, r)$ be the smallest number such that any set system with $n$ elements and more than $\sun(d, r)$ sets of cardinality $d$ must have an $r$-sunflower.
The classical bound of Erd\H{o}s and Rado~\cite{ER60} shows that $\sun(d, r) \leq d!(r - 1)^d$.
A recent breakthrough by Alweiss et al.~\cite{alweiss2019improved} proves
that $\sun(d, r) \leq  (\lg d)^d (r \cdot \lg \lg d)^{O(d)}$.

\BC\corl{sunflower}
Let $\m F$ be a family of sets over some universe, where every set has size at most $d$. If $|\m F|> (d+1) \cdot \sun(d, r)$, then $\m F$ contains an $r$-sunflower.
\EC
\BP
Group the sets in $\m F$ by their sizes, which range from $0$ to $d$. For some $d'\in[0,d]$, there are more than
\[\f1{d+1} \cdot \big((d+1) \,\sun(d, r) \big) \ge \sun(d', r) \]
sets of size exactly $d'$, since $\sun(d, r)$ is monotone in $d$. The result follows from applying the definition of $\sun(d', r)$ on the sets in $\m F$ of size $d'$.
\EP

For our applications for cuts, we want a sunflower with nonempty core. 
In this case, the bound must depend on the size of the universe $n$, since the set system with $n$ singleton sets does not contain a sunflower with nonempty core.
The following lemma proves that the above bound, multiplied by $\approx nd$, can
guarantee a sunflower with nonempty core.

\BL\leml{sunflower}
Let $\m F$ be a family of sets over a universe of $n$ elements, where every set has size at most $d$. If $|\m F|>(d + 2)\, \sun(d, r) n$, then $\m F$ contains an $r$-sunflower with nonempty core.
\EL
\BP
We prove the contrapositive: suppose that $\m F$ does not have an $r$-sunflower with nonempty core.
For each element $v\in U$, consider the set $\m F_v:=\{F\in\m F:F\ni v\}$. If there exists an $r$-sunflower in $\m F_v$ for any $v\in U$, then this sunflower has a nonempty core (since the core contains $v$), contradicting our assumption. Therefore, by \cor{sunflower},
$|\m F_v|\le (d+1) \cdot \sun(d, r)$ for each $v\in U$. Every set in $\m F$ is included in some $\m F_v$ except possibly $\emptyset$, so
\[ |\m F| \le (d+1) \,\sun(d, r) \cdot n + 1 \le (d + 2)\, \sun(d, r) n  ,\]
proving the contrapositive.
\EP

Additionally we want multiple sunflowers, each with distinct, nonempty nonempty core. Note that the sunflower cores may intersect, even though they are distinct. The following lemma shows we can also achieve this. 

\BL\leml{sunflower-nonempty}
Let $\m F$ be a family of sets over a universe of $n$ elements, where every set has size at most $d$. If $|\m F|>s (d + 2) \sun(d, r) n$, then $\m F$ contains $s$ many $r$-sunflowers, each with distinct, nonempty cores.
\EL
\BP
We iteratively construct $s$ sunflowers with distinct cores.
Initialize $\m F':=\m F$, and on each iteration, consider a maximal set
$C$ such that there exists an $r$-sunflower in $\m F'$ with core
$C$. Inductively we ensure that such a set $C \neq \emptyset$ exists;
this holds for the base case by \lem{sunflower}. 

Moreover, we claim that the set $\m F'_C:=\{F\in\m F':F\supseteq C\}$
has size at most $(d + 2) \sun(d, r) n$. Indeed, if not, then
applying~\lem{sunflower} on the set system $\{F\sm C:F\in\m
F',F\supseteq C\}$ (which has the same cardinality as $\m F'$), we
obtain an $r$-sunflower with sets $S_1,\lds,S_r$ and nonempty core
$C'$. Then, the sets $S_1\cup C,\lds,S_r\cup C \in \m F'$ form an
$r$-sunflower with core $C\cup C'$, contradicting the maximality of the
set $C$.

We now remove the sets in $\m F'_C$ from $\m F'$ (i.e., update $\m
F'\gets\m F'\sm\m F'_C$). Now the core on any subsequent iteration
cannot be $C$, since we have removed all the sets that contained $C$. The size of $\m F'$ drops by at most $|\m F'_C|\le (d + 2)\, \sun(d, r) n$ each iteration, so if $|\m F|>s (d + 2)\, \sun(d, r) n$ to begin with, then we can proceed for $s$ iterations, obtaining $s$ many $r$-sunflowers with distinct, nonempty cores.
\EP

\subsection{Removing the Size Restriction: Venn Diagrams}

The above sunflower lemmas proved that a sunflower-free set system
$\m F$ must have few sets, as long as each set in the system has bounded
size.  The following lemma replaces the assumption on the bounded size
by the assumptions that (a) every $k$ sets in the system have small
number of occupied regions in their Venn diagram, and (b) the set system
of the complements of the sets do not contain many sunflowers either.

To make this formal, we introduce some notation.  Given $k$ sets
$F_1, \dots, F_k$, we denote their {\em Venn diagram} by
$\Venn(F_1, \lds, F_k)$.  An {\em atom} denotes a nonempty region of the
diagram.  Formally, an atom is a nonempty set that can be expressed as
$G_1 \cap \lds \cap G_k$, where for each $i$, the set $G_i$ is either
the set $F_i$, or its complement $\ol{F_i}$.
% for every $i$,
% where $\ol{F_i}$ denotes the complement of $F_i$. 
Also, let $\ol{\m F} := \{ \ol{F} : F \in \m F \}$ be the collection of
complements of the sets in $\calF$.

\begin{figure}\centering
\includegraphics{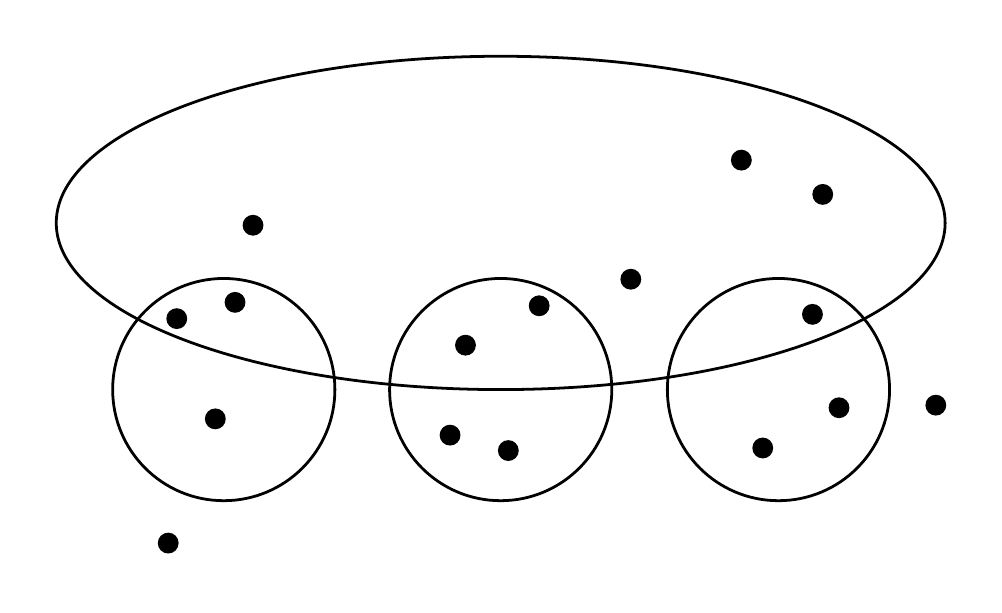}
\caption{The Venn diagram above has eight atoms.}
\end{figure}
%\elnote{(recycled) Figure?} 

\BL\leml{ex}
Let $\m F$ be a set system on $n$ elements satisfying the following:
  \BE
  \im[i.] For every $k$ sets $F_1,\lds,F_k\in\m F$, their Venn diagram $\Venn(F_1,\lds,F_k)$ has less than $2k$ atoms.
  \im[ii.] Each of $\m F$ and $\ol{\m F}$ does not contain $s$ many $r$-sunflowers, each with distinct, nonempty cores.
  \EE
  Then, $|\m F| \le 10s \cdot k(5k+2) \cdot \sun(5k, r) \cdot n$.
\EL

\BP
For fixed $r, k, s$, let $\ex(n)$ ($\ex$ for extremal) be the maximum size of a set $\m F$ on $n$ elements satisfying conditions~(i)~and~(ii). We prove by induction on $n$ that
\[ \ex(n)\le 10 s \cdot k(5k+2) \cdot \sun(5k, r) \cdot \max\{1,n-4k\},\]
with the base cases $n\le5k$.

\emph{Base case: $n\le5k$.}
In this case, each set has size at most $n \leq 5k$, so using
\lem{sunflower-nonempty}, so the number of sets in $\calF$ is at most
\[ s(5k+2) \sun(5k,r) \cd 5k \le 10 s \cdot k(5k+2) \cdot \sun(5k, r) \cdot \max\{1,n-4k\} =
  \ex(n).\]

\emph{Inductive step: $n>5k$.} First, suppose that every set $F\in\m F$
satisfies either $|F|\le5k$ or $|F|\ge n-5k$. By
\lem{sunflower-nonempty} on $\m F$ and $\ol{\m F}$ respectively, there
are at most $s (5k + 2) \cdot \sun(5k, r) \cdot n$ many sets of size at
most $5k$, and also at most $s (5k + 2) \cdot \sun(5k, r) \cdot n$ many
sets of size at least $n-5k$. Applying the bound $n\le5(n-4k)$ and using
that $k \geq 1$, we obtain
\[ |\m F| \le 2s (5k + 2) \cdot \sun(5k, r) \cdot n \le 2 s \cdot k(5k+2) \cdot \sun(5k, r) \cdot 5\max\{1,n-4k\} = \ex(n),\]
as desired.

\begin{figure}\centering
\includegraphics[scale=.5]{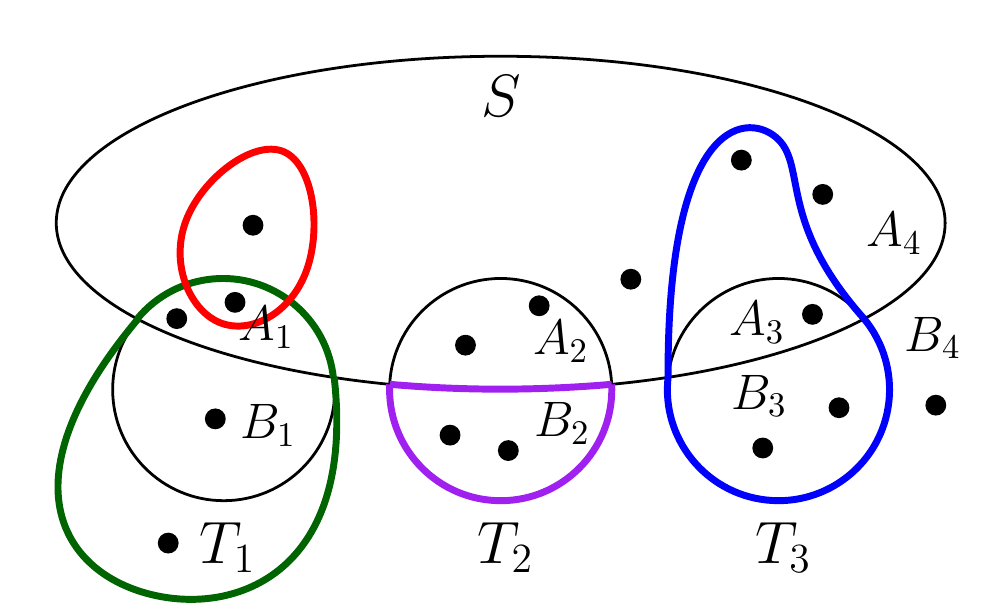}
\includegraphics[scale=.5]{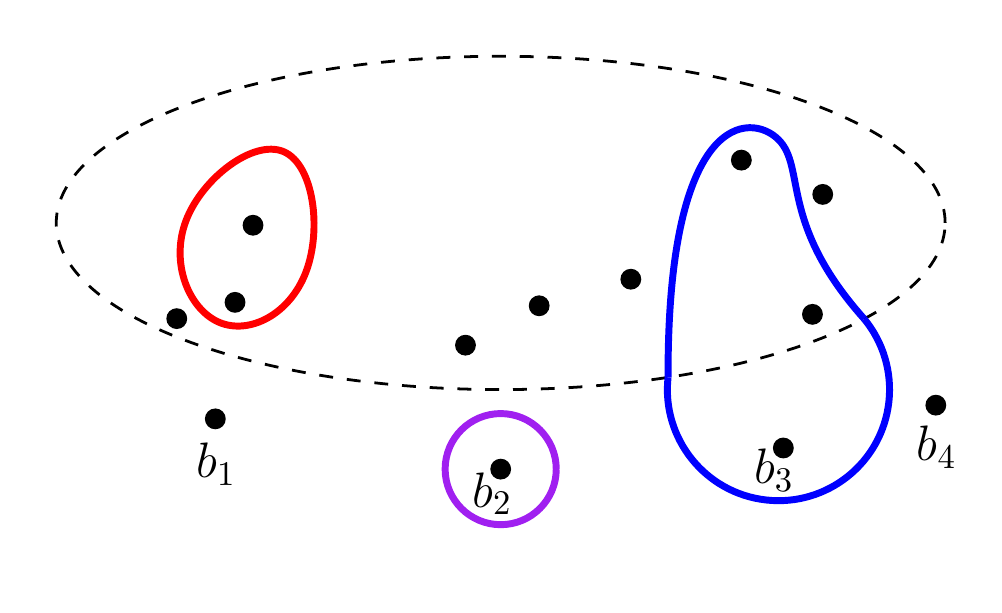}
\includegraphics[scale=.5]{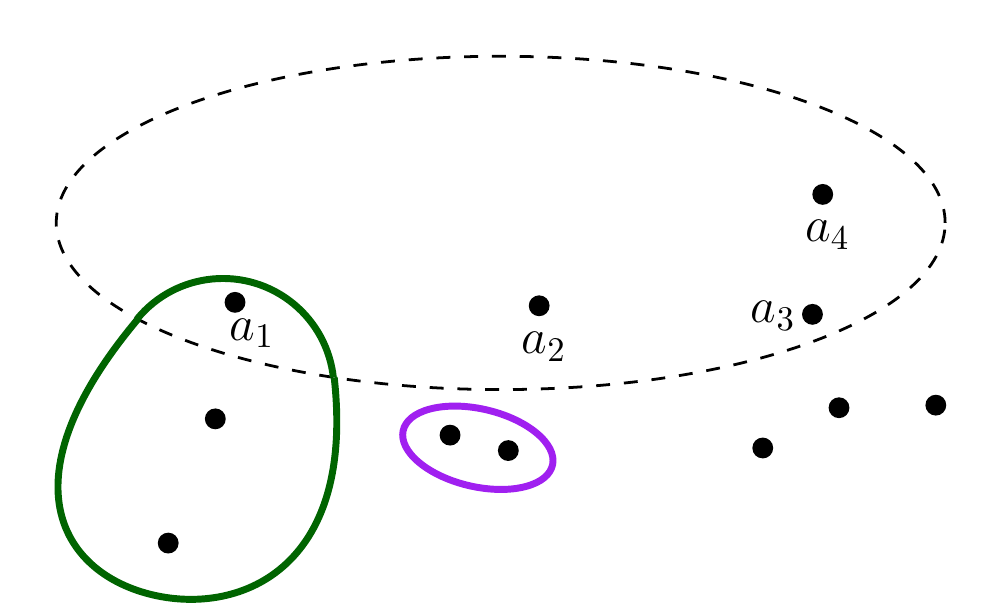}
\caption{Construction of the set systems $(X,\m F_a)$ (middle) and $(Y,\m F_b)$ (right) given the set system on the left and $S,T_1,T_2,T_3$. The purple set can be added to either $(X_,\m F_a)$ or $(Y,\m F_b)$.}
\label{fig:ex}
\end{figure}

Otherwise, there exists a set $S$ with $5k<|S|<n-5k$. For
$i=1,2,\lds,k-1$, while there exists a set $T_i\in\m F$ such that the
Venn diagram $\Venn(S,T_1,T_2,\lds,T_i)$ on the $i+1$ sets contains at
least $2(i+1)$ atoms, choose an arbitrary such set $T_i$. Suppose this
process continues until the index $i$ reaches value $\el\in[k-1]$. If
$\el=k-1$, then $\Venn(S,T_1,\lds,T_\ell)$ is composed of $k$ sets and
has at least $2k$ atoms, which cannot happen by assumption. Therefore,
$\el<k-1$. We say that a set $F$ \emph{cuts} another set $F'$ if both
the regions $F \cap F'$ and $F' \setminus F$ are non-empty. By our
stopping condition, every set $F \in \calF$ cuts at most one atom in
$\Venn(S,T_1,\lds,T_\el)$; indeed, if a set $F$ cuts two atoms or more,
we would have added it as $T_{\el+1}$ and continued.

\newcommand{\Uin}{E_a}
\newcommand{\Uout}{E_b}

Let the atoms of $\Venn(S,T_1,\lds,T_\el)$ inside $S$ be $A_1,\lds,A_i$,
so that $A_1 \cup \ldots \cup A_i = S$; and let the atoms outside $S$ be
$B_1,\lds,B_j$, so that
$B_1 \cup \ldots \cup B_j = U \setminus
S$. Define two new collections of elements $\Uin :=\{a_1,\lds,a_i\}$ and
$\Uout :=\{b_1,\lds,b_j\}$, and define $X :=S\cup \Uout$ and
$Y:=(U\sm S)\cup \Uin$. We build two set systems $(X,\m F_a)$ and
$(Y,\m F_b)$ as follows (see \Cref{fig:ex}). Initialize $\m F_a=\m F_b:=\emptyset$;
for each set $F\in\m F$, we have three cases: \BE \im If $F$ cuts an
atom $A_h$ inside $S$, then add the set
$(F\cap S)\cup \{b_h \mid h\in[j],F\supseteq B_h \}$ into $\m F_a$.  \im
Else, if $F$ cuts an atom $B_h$ outside $S$, then add the set
$(F\sm S)\cup \{a_h\mid h\in[i], F\supseteq A_h \}$ into $\m F_b$.  \im
Else, $F$ does not cut any atom. Execute either step~(1) or step~(2).
\EE

Here's another equivalent way to look at this process.
For $\m F_a$, we can think taking the set system $(U,\m F)$, removing
the sets that cut an atom outside $S$, and then contracting the atoms
$B_1,\lds,B_j$ into $b_1,\lds,b_j$, respectively. We can also think of
$\m F_b$ analogously, by throwing away the sets that cut atoms inside
$S$, and then contacting atoms $A_1, \ldots, A_i$.  Through this contraction viewpoint, it is clear
that if the set system $(U,\m F)$ satisfy conditions~(i)~and~(ii), then
so do the set systems $(X,\m F_a)$ and $(Y,\m F_b)$. Moreover, since
$5k<|S|<n-5k$, we have
\[|X| = |S|+|\Uout| \le |S|+2k\le (n-5k)+2k<|U|\]
and 
\[|Y|=|V\sm S|+|\Uin|\le n-|S|+2k\le(n-5k)+2k<|U|,\]
 so we can apply induction on $n$, obtaining
\begin{align*}
|\m F|&=|\m F_a|+|\m F_b|
\\ &\le \ex(n-|S|+2k) + \ex(|S|+2k)
\\&\le 10 s \cdot k(5k+2) \cdot \sun(5k, r) \cdot \Big(
  \max\{1,(n-|S|+2k)-4k\} + \max\{1,(|S|+2k)-4k\} \Big)
\\&= 10 s \cdot k(5k+2) \cdot \sun(5k, r) \cdot \Big(
  (n-|S|+2k)-4k + (|S|+2k)-4k \Big) 
\\&= 10 s \cdot k(5k+2) \cdot \sun(5k, r) \cdot (n-4k) = \ex(n),
%\\&= 12(5k)^2(5k)!(r-1)^{5k}\max\{1,n-4k\},
\end{align*}
completing the induction.
\EP

Later when we apply the above lemma to $k$-cut, the number of atoms becomes $k$, so it is sufficient for even $k$. 
For odd $k$, we can slightly strengthen \lem{ex} as follows.

\BC\corl{ex}
Let $\m F$ be a set system on $n$ elements satisfying the following:
  \BE
  \im[i.]  There do not exist sets $S_1,\lds,S_{k-1}$ such that $\Venn(S_1,\lds,S_{k-1})$ has at least $2(k-1)+1$ atoms.
  \im[i'.]  There do not exist sets $S_1,\lds,S_k$ such that $\Venn(S_1,\lds,S_{k-1})$ has exactly $2(k-1)$ atoms, and the set $S_k$ cuts at least two atoms in $\Venn(S_1,\lds,S_{k-1})$.
%  \im[i.]For every $k$ sets $F_1,\lds,F_k\in\m F$, their Venn diagram $\Venn(F_1,\lds,F_k)$ has less than $(2k + 1)$ atoms.
%  \im[ii.] There do not exist sets $F_1,\lds,F_k$ such that $\Venn(F_1,\lds,F_{k-1})$ has at least $2(k-1)$ atoms.
  \im[ii.] Each of $\m F$ and $\ol{\m F}$ does not contain $s$ many $r$-sunflowers, each with distinct, nonempty cores.
  \EE
Then, $|\m F| \le 10s \cdot k(5k+2) \cdot \sun(5k, r) \cdot n$.
%Suppose $|\m F|>10s(5k+2)!(s-1)^{5k}\max\{1,n-4k\}$, and that each of $\m F$ and $\ol{\m F}$ does not contain $s$ many $r$-sunflowers, each with distinct, nonempty cores. Then, there exist sets $S_1,\lds,S_k$ such that $\Venn(S_1,\lds,S_{k-1})$ has at least $2(k-1)$ atoms, and the set $S_k$ cuts at least two atoms in $\Venn(S_1,\lds,S_{k-1})$.
\EC
\BP
The proof is identical; the only additional observation is that when we
iteratively construct $S,T_1,\lds,T_\el$ for $\el\le k-1$, observe that
the set $T_{\ell}$ cuts at least two atoms of $\Venn(S,T_1,\lds,T_{\el-1})$
by construction. In particular, if the construction continued until $\el=k-1$, then %the sets $S,T_1,\lds,T_{k-1}$ would violate condition (ii) with $T_{k-1}$ as $S_k$ in the condition. 
either the sets $S,T_1,\lds,T_{k-2}$ violate condition (i), or the sets $S,T_1,\lds,T_{k-1}$ violate condition (i'). 
Therefore, every time we carry out this process, we must stop at $\el<k-1$.
When we stop, the condition (i), though it is slightly more relaxed than the condition (i) of \lem{ex}, still ensures that $|E_a|, |E_b| \leq 2k$, so the same inductive argument works.
%\elnote{Changed; see you are fine with it.}
%\alert{I don't understand this explanation: do you mean, if we run the
%  same process, then we would continue as long as the new set cuts at
%  least two atoms? And by the assumption (i) we would stop at $\el < k-1$?} \jlnote{fixed?}
\EP

\subsection{Relating Cuts and Sunflowers}
Recall that $\lk$ is the size of the minimum $k$-cut divided by $k$, and $\gamma \in [1, 2)$ is a fixed parameter.
In this section, we use the previous tools for sunflowers to bound the number of small cuts (of size $\leq \gamma \lk$) in a graph. 
First, the following lemma, independent of sunflowers, shows that there cannot be many tiny cuts (of size $< \lk$) in a graph. 

\BL\leml{small}
There are at most $2^{k-1}$ many cuts with weight less than $\lk$.
\EL
\BP
Suppose, otherwise, that there are more than $2^{k-1}$ sets; let $\m S$ be the collection of these sets. We will iteratively construct a $k$-cut of size less than $k\lk = \la_k$ contradicting the definition of $\la_k$, the size of the minimum $k$-cut.

Begin with an arbitrary set $S_1\in\m S$, and while $\Venn(S_1,\lds,S_{i-1})$ has less than $k$ components, choose an arbitrary set $S_i\in\m S$ such that $\Venn(S_1,\lds,S_i)$ has at least one more component than $\Venn(S_1,\lds,S_{i-1})$. We show that such a set $S_i$ always exists. Let $A_1,\lds,A_\el$ be the atoms of $\Venn(S_1,\lds,S_{i-1})$; the only sets $T\in\m S$ such that $\Venn(S_1,\lds,S_{i-1},T)$ has the same number of components as $\Venn(S_1,\lds,S_{i-1})$ are sets of the form $\bigcup_{i\in I}A_i$ for some subset $I\s[\el]$. Since there are at most $2^{\el}\le2^{k-1}$ such sets and $|\m S|>2^{k-1}$, a satisfying set $S_i$ always exists.

At the end, we have at most $k-1$ sets $S_1,\lds,S_i$ such that $\Venn(S_1,\lds,S_i)$ has at least $k$ components. Therefore, the edge set $\pt S_1\cup\cds\cup\pt S_i$ is a $k$-cut, and it has weight less than $i\lk<k\lk=\la_k$, achieving the desired contradiction.
\EP

Finally, the following lemma proves that many sunflowers consisting of cuts of size $\leq \gamma \lk$ will lead to a better $k$-cut than $k \lk$, 
leading to contradiction.  %\agnote{If we could somehow remove the need to have $r+k-2$ in
%  \lem{cuts-sunflower}, and instead just have $r$ there, that would be
%  a bit nicer? Maybe give us $\exp{O(k^2)}$, which is something we've
%  got in several previous algorithms. } 

\BL\leml{cuts-sunflower}
Fix a constant $1\le\g<2$, and
let $\m F$ be the family of sets $\{S\s V : w(\pt S)\le\g\lk\}$. Then, for any $r>\f{2\g}{2-\g}+1$, both $\m F$ and $\ol{\m F}$ do not contain $2^k$ many $(r+k-2)$-sunflowers with distinct, nonempty cores. 
\EL
\BP
Since $w(\pt S)\le\g\lk \iff w(\pt(V\sm S))\le\g\lk$, we have $\m F=\ol{\m F}$, so it suffices to only consider $\m F$. Suppose, otherwise, that there are $2^k$ many $(r+k-2)$-sunflowers with distinct, nonempty cores. Let $\Fsmall := \{ S \mid \emptyset\subsetneq S\subsetneq V,\, w(\partial S) <
\lk\}$, so that \lem{small} implies that $|\Fsmall|\le2^{k-1}$. Then, there must exist at least one sunflower in this collection whose core does not belong to $\Fsmall$. 
Let $S_1,\lds,S_{r+k-2}\in\m F$ be the sets of this $(r+k-2)$-sunflower with petals $P_i:=S_i\sm\bigcup_{j\ne r}S_j$ and nonempty core $C:=\bigcap_iS_i \notin \Fsmall$. Since the petals $P_i$ are disjoint, at most $k-2$ of them are in $\Fsmall$, since otherwise, we get $k-1$ disjoint sets in $\Fsmall$ which together form a $k$-cut with weight less than $(k-1)\lk<\la_k$. Therefore, without loss of generality (by reordering the sets $S_i$), assume that $P_1,\lds,P_r\notin\Fsmall$. Since $C$ and $P_1,\lds,P_r$ are all cuts in the graph (in particular, $\emptyset\ne C\ne V$ and $\emptyset\ne P_i\ne V$) and are not in $\Fsmall$, we have $w(\pt C)\ge\lk$ and $w(\pt P_i)\ge\lk$ for each $i\in[r]$. For each $i\in[r]$, we have 
\begin{align*}
\g\lk\ge w(\pt S_i)=w(\pt(C\cup P_i)) = w(\pt C)+w(\pt P_i)-2w(E[C,P_i])\ge2\lk-2w(E[C,P_i]),
\end{align*}
so $w(E[C,P_i]) \ge (2-\g)\lk/2$.
Now observe that the edges in $E[C,P_i]$ for $i=2,\lds,r$ are included in $\pt(C\cup P_1)$. It follows that
\[ \g\lk \ge w(\pt (C\cup P_1)) \ge w(E[C,P_2] \cup \cds \cup E[C,P_r]) = \sum_{i=2}^rw(E[C,P_i])\ge(r-1)\f{(2-\g)\lk}2 ,\]
so $r-1\le\f{2\g}{2-\g}$, contradicting the assumption that $r>\f{2\g}{2-\g}+1$.
\EP

%\subsection{Putting it together}

We are finally ready to prove \thm{ex-cuts}, which we restate here for convenience.

\ExtThm*

% \BT[Restatement of \thm{ex-cuts}]
% For any $\g<2$, there are $(\f k{2-\g})^{O(k)}n$ many cuts with weight less than $\g\lk$.
% \ET
\BP
Let $\m F$ be the set of such cuts, and let $k':=\lc k/2\rc$. We first show that if $k$ is even, then condition~(i) of \lem{ex} is satisfied when the parameter $k$ in the lemma is $k'$ instead, and if $k$ is odd, then conditions~(i)~and~(i') of \cor{ex} are satisfied, again with $k'$ for the parameter $k$. Then, we show that condition~(ii) of both \lem{ex} and \cor{ex} are satisfied for parameters $s$ and $r$ that we choose later.

First, consider the case when $k$ is even. Suppose, otherwise, that condition~(i) of \lem{ex} is false: there are sets $S_1,\lds,S_{k'}$ such that $\Venn(S_1,\lds,S_{k'})$ has at least $2k'=k$ atoms. Then, $\pt S_1\cup\cds\cup\pt S_{k'}$ is a $k$-cut with weight $k'\cd \g\lk < \f k2\cd2\lk = \la_k$, contradicting the definition of $\la_k$, the minimum $k$-cut. 

Now consider the case when $k$ is odd. If condition~(i) of \cor{ex} is false, then there are sets $S_1,\lds,S_{k'-1}$ such that $\Venn(S_1,\lds,S_{k'}$ has at least $2(k'-1)-1=k$ atoms. Then, $\pt S_1\cup\cds\cup\pt S_{k'-1}$ is a $k$-cut with weight $(k'-1)\cd \g\lk < \f k2\cd2\lk = \la_k$, contradicting the definition of $\la_k$, the minimum $k$-cut.
Otherwise, if condition~(i') of \cor{ex} is false, then the set of edges $\pt S_1\cup\cds\cup\pt S_{k'-1}$ is a $(k-1)$-cut with weight $(k'-1)\cd\g\lk < \f{k-1}2\cd2\lk=(k-1)\lk$. Let $A_1,\lds,A_\el\s V$ be the atoms in $\Venn(S_1,\lds,S_{k'-1})$ that are cut by $S_{k'}$, with $\el\ge2$ by assumption. Since $\pt S_{k'} \cap E[A_i]$ are disjoint for $i\in[\el]$, there exists one atom $A_i$ such that 

\[ w(\pt S_{k'}\cap E[A_i]) \le \f1\el w(\pt S_{k'}) \le \f1\el\cd\g\lk < \f12\cd2\lk=\lk .\]
Thus, $\pt S_1\cup\cds\cup\pt S_{k'-1}\cup (\pt S_{k'}\cap E[A_i])$ is a $k$-cut with weight less than $(k-1)\lk+\lk=\lk$, a contradiction. Thus, conditions~(i)~and~(i') of \cor{ex} are satisfied.

Now, fix the parameters $s:=2^k$, $r:=\lc\f{2\g}{2-\g}+2\rc$, $r':=r+k-2$, and $k':=\lc k/2\rc$. By \lem{cuts-sunflower}, both $\m F$ and $\ol{\m F}$ do not contain $s$ many $r'$-sunflowers with distinct, nonempty cores, fulfilling condition~(ii) of \lem{ex} and \cor{ex} (with $r'$ in place of $r$). Therefore, by \lem{ex} or \cor{ex} when $k$ is even or odd respectively, 
and using $\sun(d, r) \leq  (\lg d)^d (r \cdot \lg \lg d)^{O(d)} \leq (r \lg d \lg \lg d)^{O(d)}$~\cite{alweiss2019improved}, %\jlnote{Verified that it's indeed log base 2 in ALWZ}
\begin{align*} |\m F|&\le 
10s \cdot k'(5k'+2) \cdot \sun(5k', r') \cdot n
%10 s \cdot 2^{5k'} \cdot \sun(5k', r') \cdot n
\le 2^{O(k)} \cdot (r' \lg k' \lg \lg k')^{O(k')}n \\
& =  2^{O(k)} \cdot \lp \lp  \lc\f{2\g}{2-\g}\rc + k \rp \lg k' \lg \lg k' \rp^{O(k')} \cdot n
\leq \lp \max\lp \frac{1}{2 - \gamma}, k \rp \rp^{O(k)} \cdot n
%&= \lp \left\lc\f k2\right\rc\lp\left\lc\f{2\g}{2-\g}+2\right\rc+k-2\rp\rp^{O(k)}n = \lp\f k{2-\g}\rp^{O(k)}n ,
\end{align*}
as desired. As an aside, using the classical Erd\H{o}s-Rado
bound~\cite{ER60} of $\sun(d, r) \leq d!(r - 1)^d$ gives the same result,
up to a constant in the $O(k)$ exponent, as does the conjectured optimal
bound of $\sun(d,f) \leq (Cr)^d$.
\EP

%%% Local Variables:
%%% mode: latex
%%% TeX-master: "main"
%%% End:

{\small
  \bibliographystyle{alpha}
\bibliography{refs}

\begin{thebibliography}{ALWZ19}

\bibitem[ALWZ19]{alweiss2019improved}
Ryan Alweiss, Shachar Lovett, Kewen Wu, and Jiapeng Zhang.
\newblock Improved bounds for the sunflower lemma.
\newblock {\em arXiv preprint arXiv:1908.08483}, 2019.

\bibitem[CQX18]{chekuri2018lp}
Chandra Chekuri, Kent Quanrud, and Chao Xu.
\newblock {LP} relaxation and tree packing for minimum k-cuts.
\newblock In {\em 2nd Symposium on Simplicity in Algorithms (SOSA 2019)}.
  Schloss Dagstuhl-Leibniz-Zentrum fuer Informatik, 2018.

\bibitem[ER60]{ER60}
P.~Erd\H{o}s and R.~Rado.
\newblock Intersection theorems for systems of sets.
\newblock {\em J. London Math. Soc.}, 35:85--90, 1960.

\bibitem[Fre75]{freedman1975tail}
David~A Freedman.
\newblock On tail probabilities for martingales.
\newblock {\em the Annals of Probability}, 3(1):100--118, 1975.

\bibitem[GH94]{GH94}
Olivier Goldschmidt and Dorit~S. Hochbaum.
\newblock A polynomial algorithm for the {$k$}-cut problem for fixed {$k$}.
\newblock {\em Math. Oper. Res.}, 19(1):24--37, 1994.

\bibitem[GLL18]{GLL18focs}
Anupam Gupta, Euiwoong Lee, and Jason Li.
\newblock Faster exact and approximate algorithms for $k$-cut.
\newblock In {\em Foundations of Computer Science (FOCS), 2018 IEEE 59th Annual
  Symposium on}, 2018.

\bibitem[GLL19]{GLL19}
Anupam Gupta, Euiwoong Lee, and Jason Li.
\newblock The number of minimum k-cuts: improving the karger-stein bound.
\newblock In {\em Proceedings of the 51st Annual ACM SIGACT Symposium on Theory
  of Computing}, pages 229--240. ACM, 2019.

\bibitem[KS96]{KS96}
David~R. Karger and Clifford Stein.
\newblock A new approach to the minimum cut problem.
\newblock {\em Journal of the ACM (JACM)}, 43(4):601--640, 1996.

\bibitem[KYN07]{KYN06}
Yoko Kamidoi, Noriyoshi Yoshida, and Hiroshi Nagamochi.
\newblock A deterministic algorithm for finding all minimum {$k$}-way cuts.
\newblock {\em SIAM J. Comput.}, 36(5):1329--1341, 2006/07.

\bibitem[Li19]{Li19focs}
Jason Li.
\newblock Faster minimum k-cut of a simple graph.
\newblock In {\em Foundations of Computer Science (FOCS), 2019 IEEE 60th Annual
  Symposium on}, 2019.

\bibitem[NI92]{NI92}
Hiroshi Nagamochi and Toshihide Ibaraki.
\newblock Computing edge-connectivity in multigraphs and capacitated graphs.
\newblock {\em SIAM J. Discrete Math.}, 5(1):54--66, 1992.

\bibitem[Tho08]{Thorup08}
Mikkel Thorup.
\newblock Minimum $k$-way cuts via deterministic greedy tree packing.
\newblock In {\em Proceedings of the fortieth annual ACM symposium on Theory of
  computing}, pages 159--166. ACM, 2008.

\end{thebibliography}
}

\appendix
\section{Omitted Proofs}
\BP[Proof of \clm{diffeq}]
We have
\[ \tilde f(0) = \f1{1-A}n(1 - A) = n = f(0) \]
and 
\begin{align*}
\tilde f'(t) &= \f1{1-A}n(-\f\g2e^{-(\g/2)t} + Ae^{-t})
\\&= \f1{1-A}n\lp-\f\g2e^{-(\g/2)t} + \f\g2Ae^{-t} + \lp A-\f\g2A\rp e^{-t}\rp \\&= -\f\g2\cd\f1{1-A}n(e^{-(\g/2)t}-Ae^{-t}) + \f{A-(\g/2)A}{1-A}e^{-t}n
\\&= -\f\g2\tilde f(t) + \f{A-(\g/2)A}{1-A}e^{-t}n.
\\&= -\f\g2\tilde f(t) + Be^{-t}n ,
\end{align*} 
where the last equality holds because
\begin{align*}
A = \f{B}{B+1-\g/2} &\iff BA+A-\f\g2A = B 
\\&\iff A-\f\g2A= B-BA=B(1-A) 
\\&\iff \f{A-(\g/2)A}{1-A} = B.
\end{align*}
Since $\tilde f(0)=f(0)$ and $\tilde f(t)$ satisfies \eqn{f'} with equality, we have $\tilde f(t)\ge f(t)$ for all $t\ge0$.
\EP

\iffalse
\BP[Proof of \clm{recursion-xi}]
We shall assume the two inequalities
\begin{gather}
\lg  (5\cd2^{-i}\xx_0) \le 2^{-i}\xx_0 \eqnl{11}
\end{gather}
and
\begin{gather}
Ck\lg k\le \f12 C^2k\lg k -4 , \eqnl{12}
\end{gather}
which indeed hold for large enough $C$ and all $i \le \lg_2\xx_0-C$.  The inductive statement \eqn{ind} is clearly true for $i=0$, and for $i>0$, we have
\begin{align*}
\xx_i &\stackrel{\eqn{xi}}\le Ck\lg k + 2\lg \xx_{i-1} + \f12 \xx_{i-1}
\\&\le Ck\lg k+2\lg\left(C^2k\lg k+4(\lg  (5\cd2^{-(i-1)}\xx_0))+ 2^{-(i-1)}\xx_0\right) + \f12 \xx_{i-1}
\\&\le Ck\lg k+2\lg(C^2k\lg k) + 2\lg \left(4(\lg  (5\cd2^{-(i-1)}\xx_0))+ 2^{-(i-1)}\xx_0\right)+ \f12 \xx_{i-1}
\\&\stackrel{\eqn{11}}\le Ck\lg k+2\lg(C^2k\lg k) + 2\lg \left(4(2^{-(i-1)}\xx_0)+ 2^{-(i-1)}\xx_0\right)+ \f12 \xx_{i-1}
\\&= Ck\lg k+2\lg(C^2k\lg k) + 2\lg \left(5\cd2^{-(i-1)}\xx_0\right)+ \f12 \xx_{i-1}
\\&\stackrel{\eqn{12}}\le \lp\f12 C^2k\lg k -4 \rp +2\lg \left(5\cd2^{-(i-1)}\xx_0\right)+ \f12\xx_{i-1}
\\&\stackrel{\eqn{ind}}\le  \lp\f12 C^2k\lg k -4 \rp  +2\lg \left(5\cd2^{-(i-1)}\xx_0\right)+ \f12\left(C^2k\lg k+4(\lg(5\cd2^{-(i-1)}\xx_0))+2^{-(i-1)}\xx_0\right)
\\&=   C^2k\lg k  +4\lg \left(5\cd2^{-(i-1)}\xx_0\right)-4+2^{-i}\xx_0 
\\&=   C^2k\lg k  +4\lg \left(5\cd2^{-i}\xx_0\right)+2^{-i}\xx_0 ,
\end{align*}
completing the induction. 
\EP
\fi

%%% Local Variables:
%%% mode: latex
%%% TeX-master: "main"
%%% End:

\end{document}